\documentclass[12pt]{iopart}
\usepackage{graphicx}
\usepackage{xcolor}
\usepackage{subfig}
\usepackage{floatrow} 
\usepackage{url}
\usepackage{rotating}
\usepackage{lscape}
\usepackage{bm}
\usepackage[numbers,sort&compress]{natbib}

\newcommand{\eqref}[1]{(\ref{#1})}

\begin{document}
\floatsetup[figure]{style=plain,subcapbesideposition=top}

\title[Model reconstruction for oscillator networks]{Model reconstruction from temporal data for coupled oscillator networks}

\author{Mark J. Panaggio$^{1}$, Maria-Veronica Ciocanel$^{2}$, Lauren Lazarus$^{3}$, Chad M. Topaz$^{4}$, Bin Xu$^{5}$}

\address{$^{1}$Department of Mathematics, Hillsdale College, Hillsdale, MI 49242, USA}
\address{$^2$Mathematical Biosciences Institute, The Ohio State University, Columbus, OH 43210, USA}
\address{$^{3}$Department of Mathematics, Trinity College, Hartford, CT 06106, USA}
\address{$^{4}$Department of Mathematics and Statistics, Williams College, Williamstown, MA 01267, USA}
\address{$^{5}$Department of Applied and Computational Mathematics and Statistics, University of Notre Dame, Notre Dame, IN 46556, USA}
\eads{\mailto{mpanaggio@hillsdale.edu}}

\begin{abstract}
In a complex system, the interactions between individual agents often lead to emergent collective behavior like spontaneous synchronization,  swarming,  and  pattern  formation.   The  topology  of  the  network  of  interactions  can  have  a  dramatic influence over those dynamics.  In many studies, researchers start with a specific model for both the intrinsic dynamics of each agent and the interaction network, and attempt to learn about the dynamics that can be observed in the model.  Here we consider the inverse problem: given the dynamics of a system, can one learn about the underlying network?  We investigate arbitrary networks of coupled phase-oscillators whose dynamics are characterized by synchronization.  We demonstrate that, given sufficient observational data on the transient evolution of each oscillator, one can use machine learning methods to reconstruct the interaction network and simultaneously identify the parameters of a model for the intrinsic dynamics of the oscillators and their coupling.  
\end{abstract}


%\submitto{\NJP}

\section{Introduction}

Nature and society brim with systems of coupled oscillators, including pacemaker cells in the heart, insulin-secreting cells in the pancreas, neural networks in the brain, fireflies that synchronize their flashing, chemical reactions, Josephson junctions, power grids, metronomes, and applause in human crowds, to name merely a few \cite{TraKas1961,StaSamWei1980,BuzDra2004,Buc1988,TinNkoSho2012,HadBeaWie1988,MotMyeAng2013,Pan2002,NedRavErz2000}. The dynamics of coupled oscillators in complex networks have been studied extensively.  In particular, networks of interacting oscillators governed by the seminal Kuramoto model \cite{Kur1975,AceBonPer2005,AreDiaKur2008} or its variants are known to exhibit a rich variety of behaviors including spontaneous synchronization, phase transitions, and pattern formation \cite{Str2000,PanAbr2015}. The connections in the network can play a pivotal role in determining these dynamics.

Given oscillators where the governing equations and network topology are known, it is fairly straightforward to explore the dynamics of the system using numerical methods and, in special cases, analytical methods. Unfortunately, in many systems, the topology of the interaction network  and the intrinsic oscillator properties can be difficult or impossible to observe directly. Imagine, for example, experimental results that track neuronal cell network or gene regulatory network activity for a large number of oscillators, giving rise to time series data. One might like to infer model information from these time series.

It is crucial to distinguish between functional network connections and structural network connections. Functional connectivity refers to the temporal correlation of the oscillators and is often directly observable. For instance, neurons that fire synchronously are functionally connected. In contrast, structural connectivity, also called network topology, refers to the underlying connections present in the network. For instance, neurons that are connected by  synapses or neurotransmitters are structurally connected. In many systems these structural connections can be difficult or impossible to observe. 

Network reconstruction is an active area of research \cite{WanCheHua2009,TaYooHol2010,ShaTim2011,ChiLaiLeu2013,ChiLaiLeu2015,LinWanYin2015,TimCas2014,TirSevBul2015,TirSevRic2015,BiaRubAnt2016,AldFioDib2017,AngTulMor2017,StoHarPor2017,Pik2018,LegLevTod2019}. In \sref{sec:inv}, we review the literature most closely related to our approach. In our work, we solve an inverse problem using observed time series data of the coupled oscillators to reconstruct both the network topology, i.e. the structural network connections, and the intrinsic oscillator dynamics. For the remainder of this paper, when we refer to inferring or reconstructing the model, we mean both of the aforementioned components: network connections and intrinsic oscillator properties.

In this paper, our primary contribution is an algorithm that addresses certain challenges of existing methods mentioned above. One such challenge is that existing methods set up the inverse problem as a linear system that involves many unknown parameters. Consequently, a large amount of time series data is necessary. In contrast, our algorithm results in a system of equations that is smaller, and is nonlinear. We solve these using optimization tools designed for neural networks. As a result, we are able to infer the model with a much smaller amount of data. A second challenge, as mentioned in \cite{TimCas2014,ShaTim2011,Pik2018}, is that it is typically quite difficult to infer the model for networks that are synchronized. For such networks, we explore what perturbations of the synchronized state are needed to enable accurate inference of the model.

The mathematical setting of our study is oscillator networks  composed of limit-cycle oscillators. These are also known as phase oscillators, meaning they are characterized by a single phase variable defined on the circle. We explore four different choices of oscillator models, all of which fall into the general framework for uniformly coupled phase oscillators developed in \cite{Dai1992}: the classic Kuramoto model \cite{Kur1975}; a Kuramoto-like model with square-wave coupling function; the Kuramoto-Sakaguchi model \cite{SakKur1986}; and a phase-field reduction of weakly-coupled Hodgkin-Huxley oscillators \cite{HanMatMeu1993} (see \ref{app:coup_funcs}). We take all networks to be Erd\"os-R\'enyi random graphs \cite{ErdRen1960}. We attempt to infer the network topology, the intrinsic frequency of each model oscillator, and the so-called coupling function that specifies the influence of one oscillator on others.

To carry out the model reconstruction, we begin with simulated time series for the oscillator phases and estimated phase velocities. We set up an optimization problem involving the mean squared error for the predicted phase velocities generated by a system of nonlinear differential equations, which can be represented as a convolutional neural network. We then use computational methods designed for neural networks to estimate the optimal parameters, thereby inferring the adjacency matrix for network connectivity, the oscillator coupling strength, the oscillator frequencies, and the Fourier coefficients of the coupling function. This approach is effective for a variety of different networks and model parameters. More specifically, we find that:
\begin{itemize}
    
\item    Accurate inference of the coupling network, frequencies and coupling functions is possible independent of the model when the system does not synchronize or when sufficient perturbations from synchronization are permitted.
    
\item    Model reconstruction is possible with a smaller amount of data than previous approaches that set up the problem as a large system of linear equations.
    
\item    Computational methods designed for neural networks, such as mini-batch gradient-descent implemented in TensorFlow \cite{tensorflow2015-whitepaper}, can be used to solve the optimization problem associated with model reconstruction.
    
\item     Synchronizing networks can be reconstructed using random phase resets or sufficiently large phase perturbations to a small randomly-chosen set of oscillators.
    
\item    Synchronizing networks can also be reconstructed using perturbations to a sufficiently large fixed subset of oscillators.
    
\end{itemize}

The rest of this paper is organized as follows. In \sref{sec:methods} we introduce the Kuramoto model (a standard model for the dynamics of coupled oscillators), set up the inverse problem for reconstruction, and discuss approaches for solving this inverse problem. In \sref{sec:experiments} we describe a series of experiments used to test the effectiveness of these reconstruction techniques over a wide range of model parameters. In \sref{sec:results}, we present the results of the aforementioned experiments and discuss perturbation strategies for improving the reconstruction when the system synchronizes. Finally, in \sref{sec:conclusions}, we summarize our primary findings and discuss extensions to our methodology.

\section{Models and Methods} \label{sec:methods}

The Kuramoto model is a standard model for the dynamics of coupled oscillators.  We consider $N$ oscillators with phases $\theta_k \in [\,0,2\pi)$ for $k=1,2,\ldots,N$, each with an intrinsic frequency $\omega_k$. These oscillators are coupled through an interaction network with adjacency matrix $A$, with the entry $A_{kj}\in\{0,1\}$ determining whether oscillators $k$ and $j$ are coupled. The dynamics of $\theta_k$ are governed by the equation
\begin{equation}\label{eq:kuramoto}
\dot{\theta}_k = \omega_k + \sum_{j=1}^N \sigma_{kj}A_{kj} \, \Gamma(\theta_j-\theta_k).
\end{equation}
Here $\sigma_{kj}$ represents the strength of the coupling between oscillators $j$ and $k$, and $\Gamma(\theta)$ represents the coupling function. If the coupling function $\Gamma(\theta)$ is continuous and $2\pi-$periodic, and $\Gamma'(\theta)$ is piecewise continuous, then it can be represented using a uniformly convergent Fourier series 
\begin{equation}
\Gamma(\theta)=a_0+\sum_{n=1}^\infty a_n\cos(n x)+b_n\sin(n x)\,,
\end{equation} 
where the coefficients satisfy $a_n,b_n \leq M/n^2$ for some $M$. Therefore, these coefficients decay to $0$ for large $n$ and this function can be approximated to arbitrary accuracy using a truncated Fourier series
\begin{equation}\label{eq:fourier} \Gamma(\theta)=a_0+\sum_{n=1}^m \left[a_n\cos(n\theta)+b_n\sin(n\theta)\right] 
\end{equation}
with sufficiently large $m$. Note that within the model, one can set $a_0=0$ without loss of generality, using the change of variables $\omega_k+\sum_{j=1}^N\sigma_{kj}A_{kj}a_0\rightarrow \omega_k$. 

In Kuramoto's original formulation \cite{Kur1975}, the oscillators were globally coupled ($A_{kj}=1$) with coupling strengths $\sigma_{kj}=K/N$ where $K$ scales the global coupling strength.  He used the coupling function $\Gamma(\theta)=\sin(\theta)$, and showed that, for $K>K_C$ where $K_C$ is a critical value related to the width of the distribution of intrinsic frequencies, the oscillators begin to synchronize, achieving identical phase velocities with phases distributed around the population mean. Analogous results were obtained later for systems with periodic coupling functions possessing only odd harmonics \cite{Dai1992,AceBonPer2005} and arbitrary complex networks \cite{RodrPer2016}.

\subsection{Simulated data generation}
We investigate a method for inferring the parameters of the Kuramoto model for phase oscillators on random graphs with uniform coupling strengths $\sigma_{kj}=\frac{K}{N}$. In our experiments, we focus on undirected Erd\H{o}s-R\'enyi graphs, where each possible edge in the network is present with probability of $p$. It is straightforward to extend our approach to arbitrary networks with non-uniform coupling strengths, though this would likely require a larger amount of time series data. The details of our method appear in sections \ref{sec:inv} and \ref{sec:recon}.

\begin{table}[h!]
\caption{Summary of network model parameters, with the default values and value ranges considered in the parameter sweep experiments of \sref{sec:experiments}.}
\label{tab:networkparams}
\begin{tabular}{@{}l*{15}{l}}
\br
Parameter & Description & Value & Sweep values\\
\mr
$N$ & number of oscillators & 10 &$\left\{5,10,20,40\right\}$\\
$\mu$ & average intrinsic frequency & 1.0 & N/A\\[10pt]
$\sigma$ & \begin{minipage}{0.3\columnwidth} standard deviation \\of intrinsic frequencies \end{minipage}& 0.5 & $\left\{0.01,0.1,1.0\right\}$  \\[10pt]
$p$ & network connection probability & 0.5 & $\left\{0.1,0.2,\ldots,0.9\right\}$ \\
$\Gamma(x)$ & coupling function & $\sin(x)$ & See \ref{app:coup_funcs}\\
dyn\_noise & noise in system dynamics & 0 & $\left\{0,10^{-5},10^{-4},\ldots,10^{0}\right\}$ \\
\br
\end{tabular}
\end{table}
\begin{table}[h!]
\caption{Summary of algorithm parameters, with the default values and value ranges considered in the parameter sweep experiments of \sref{sec:experiments}.}
\label{tab:solutionparams}
\begin{tabular}{@{}l*{15}{l}}
\br
Parameter & Description & Value & Sweep values\\
\mr
$t_{max}$ & duration of each transient & 20 & $\left\{2,5,10,20,50\right\}$\\
$\Delta t$ & sampling time step & 0.1 & N/A\\
noise & observation noise & 0 & $\left\{0,10^{-5},10^{-4},\ldots,10^{0}\right\}$\\
$N_{res}$ & number of transients observed & 10 & $\left\{1,2,5,10,20,40\right\}$\\

\br
\end{tabular}
\end{table}
In order to test our method, we generate data according to the following procedure:

\begin{enumerate}
\item Generate an Erd\H{o}s-R\'enyi network for a fixed connection probability $p$ where the nodes represent individual oscillators with natural frequencies sampled from a normal distribution with mean $\mu$ and standard deviation $\sigma$.  

\item Use numerical integration to generate a time series for the evolution of this system for $t\in[0,t_{max}]$, starting from initial phases drawn from a uniform distribution on $[0,2\pi]$. For numerical integration, we use an explicit Runge-Kutta method of order 5(4) \cite{DormPrin1980} or a stochastic Runge-Kutta method of order 2 \cite{Honeycutt1992}.
\item Compute the phases $\theta_k(t_n)$ at times $t_n=n\Delta t$ with timestep $\Delta t$ and for $n=0,1,\ldots T$ where $T=t_{max}/\Delta t$ to obtain observations that are evenly spaced in time. 
\item Estimate the phase velocities $v_k(t_n)=\dot{\theta}_k(t_n)$ using central differencing in time with Savitsky-Golay filtering \cite{Savitzky1964}. We used a window length of $5$ with first degree polynomials. 
\item Repeat steps (ii-iv) $N_{res}$ times with different uniform random initial conditions to obtain sufficient data during the transient evolution of the system. 
\end{enumerate}

See tables \ref{tab:networkparams} and \ref{tab:solutionparams} for the network parameters and numerical solution parameters used in this procedure.

The aforementioned process is intended to produce simulated data mimicking that which experimentalists might collect when observing real world networks. Note however, that it may not be possible to control the initial phases in an experiment. We therefore evaluate the reconstruction methods for varying $N_{res}$, as well as for cases in which small perturbations are used instead of different initial conditions (see \sref{subsec:perturbations}).

\subsection{Inverse problem formulation}\label{sec:inv}
We will formulate a set of equations whose least squares solution can be used to estimate the natural frequencies $\omega_k$ of each oscillator, the adjacency matrix $A_{kj}$ of the coupling network, the coupling strength $K$, and the coupling function $\Gamma(\theta)$, from observed phases $\theta_k(t_n)$ and phase velocities $v_k(t_n)$ generated using the method outlined above. This work builds on prior work by Shandilya and Timme \cite{ShaTim2011}, where the adjacency matrix for a network of oscillators is estimated from a time series. We summarize their approach below. 

Define vectors $\bm{\theta}_j=[\theta_1(t_j),\theta_2(t_j),\ldots,\theta_N(t_j)]^T$ and $\bm{v}_j=\left[\dot{\theta}_1(t_j),\dot{\theta}_2(t_j),\ldots,\dot{\theta}_N(t_j)\right]^T$ consisting of the oscillator phases and phase velocities. When the coupling strength $K$ and coupling function $\Gamma$ are known, the phase velocity for oscillator $k$,
\begin{equation}\label{eq:phase_velocity}
\left(\bm{v}_j\right)_k= \omega_k + \frac{K}{N}\sum_{j=1}^N A_{kj} \, \Gamma(\theta_j-\theta_k),\end{equation}
is linear in the unknown parameters $\omega_k$ and $A_{kj}$.  Therefore, one can infer both the natural frequencies and adjacency matrix by solving a linear system of $N\cdot T$ equations, where $T$ is the number of time steps and each timestep $j$ provides $N$ equations of the form
\begin{equation}
L(\bm{\theta}_j)\bm{x}=\bm{v}_j.
\end{equation}
Here $L(\bm{\theta}_j)$ is an $N \times (N+N^2)$ matrix where each row is determined from \eqref{eq:phase_velocity} for a particular oscillator, and $\bm{x}$ is an $(N+N^2) \times 1$ vector with the unknown natural frequencies $\bm{\omega}$ and the entries in the adjacency matrix $A$. The number of unknowns in $A$ can be reduced to $N+N(N-1)/2$ if one assumes $A_{jj}=0$ and $A_{kj}=A_{jk}$.

Since the number of equations in the linear system is determined by the number of observed timesteps $T$ and the number of oscillators $N$, one can ensure that the system is over-determined by collecting enough observations so that $T>1+N$. One can then estimate the parameters by minimizing  a loss function such as the mean squared error,
\[E(\bm{\omega},A)=\frac{1}{T}\sum_{j=1}^T\|\bm{v}_j -\hat{\bm{v}}_j(\bm{\omega},A)\|_2^2\]
where $\bm{v}_j$ denotes the observed value and $\hat{\bm{v}}_j$ denotes the predicted value of the velocity. 

As long as the system remains far from synchronization, increasing the number of observations $T$ will provide additional linearly independent equations to aid with reconstruction.  However, as networks synchronize, i.e. $(\bm{v}_j)_k=\dot{\theta}_k\rightarrow\dot{\theta}$,  additional observations become (nearly) linearly dependent. This causes the system of equations to become highly ill-conditioned and makes numerical solutions sensitive to noise and rounding errors. As a result, it is often necessary to perturb the system away from equilibrium to ensure that a sufficient number of linearly independent observations can be collected. 

Although the approach of \cite{ShaTim2011} outlined above is effective, it is limited by the requirement that the coupling function be known a priori. In \cite{Pik2018}, Pikovsky addresses this limitation by expressing $\Gamma$ as a Fourier series so that it, too, can be estimated by solving an analogous optimization problem. However, a challenge remains. If one represents the coupling function as in  \eqref{eq:fourier}, then the system of equations is no longer linear as terms of the form $A_{kj}a_n$ and $A_{kj}b_n$ appear. Pikovsky \cite{Pik2018} circumvents this issue by defining distinct coupling functions 
\[\Gamma_{kj}(\theta)=\sum_{n=1}^m a_{kjn}\cos(n\theta)+b_{kjn}\sin(n\theta)\]
for each pair of oscillators and by setting $A_{kj}=1$. In this formulation, if oscillators $k$ and $j$ are uncoupled, then $a_{kjn}=b_{kjn}=0$ for all $n$.  This modification preserves the linearity of the system of equations and allows for the description of a more general class of networks with distinct coupling functions for each pair of oscillators.  Unfortunately, it comes at a cost: the number of unknown parameters increases dramatically from $N+N^2$ to $N+2mN^2$. Therefore, even longer observation times are required. Additionally, as with the method of \cite{ShaTim2011}, if the system synchronizes, there may be numerical difficulties when inferring parameters. Finally, the large number of free parameters makes this model particularly prone to overfitting. As before, one could assume that connections in the network are symmetric, \emph{i.e.}, that $\Gamma_{kj}(\theta)=\Gamma_{jk}(\theta)$ and that self edges are not included $\Gamma_{jj}(\theta)=0$, allowing for a reduction in the number of parameters to $N+mN(N-1)$.

We propose an alternative approach.  We use a single coupling function represented by a Fourier series as described in \eqref{eq:fourier} for all coupling terms and attempt to infer the $2m$ Fourier coefficients $\bm{a}=[a_1,a_2,\ldots,a_m]^T$ and $\bm{b}=[b_1,b_2,\ldots,b_m]^T$ along with the intrinsic frequencies $\bm{\omega}$, adjacency matrix $A$, and global coupling strength $K$.  This leads to a total $2m+N+N^2+1$ inferred parameters or $2m+N+N(N-1)/2+1$ with symmetry constraints and no self-connections. This is therefore a modest increase of $2m+1$ parameters over the case considered by Shandilya and Timme \cite{ShaTim2011} without requiring prior knowledge of the coupling function.

As mentioned previously, this leads to a set of nonlinear equations due to the appearance of terms of the form $KA_{kj}a_{n}$ and $KA_{kj}b_{n}$ in \eqref{eq:phase_velocity}. Since  $K$ always appears in products involving $a_n$ and $b_n$, these parameters are not structurally identifiable. One could set $K=1$ without loss of generality. Instead, we include $K$ as an inferred parameter and introduce penalty terms to the objective function we seek to minimize:
\begin{eqnarray}
\label{eq:penalty}
E(\bm{\omega},A,\mathbf{a},\mathbf{b})& = & \frac{1}{T}\sum_{j=1}^T\|\bm{v}_j -\hat{\bm{v}}_j(\bm{\omega},A,\bm{a},\bm{b},K)\|_2^2+\lambda_\Gamma\sum_{n=1}^\infty \left(|a_n|^2+|b_n|^2\right)
\\
&&\mbox{}+\sum_{j=1}^N\sum_{k=1}^N \lambda_A\left(|A_{kj}|^2+|1-A_{kj}|^2\right) \nonumber\\
&&\mbox{}+\sum_{j=1}^N\sum_{k=1}^N
+\lambda_{bd}\left( \min\left\{ A_{kj},0 \right\}+\min\left\{ 1-A_{kj},0 \right\}\right).\nonumber
\end{eqnarray}

The inclusion of a penalty function ensures the existence of a non-degenerate local minimum. In this objective function, the first term above represents the mean squared error. The remaining terms are penalty terms with weights $\lambda_\Gamma=0.0001$, $\lambda_A=10^{-6}$, and $\lambda_{bd}=10^5$. These hyper-parameter values were tuned initially to provide satisfactory reconstruction performance, and then kept constant in all subsequent experiments (see also \tref{tab:methodparams}). Additional tuning of these parameters for specific networks could further improve the accuracy of the model reconstruction.  The second term with $\lambda_\Gamma$ introduces $L_2$ regularization which favors smaller estimates of the parameter values $a_n$ and $b_n$. This is useful for combating overfitting and promoting sparse representations, and is analogous to using a Bayesian prior centered at 0 for the Fourier coefficients \cite{murphy2012machine}. The last term with $\lambda_{A}$ and $\lambda_{bd}$ penalizes adjacency values that are far from 0 and 1 as well as those that are negative or greater than 1 to ensure that the estimates fall within the desired range of $[0,1]$. We do not penalize $K$ and instead allow it to be as large as necessary to counteract the parameter shrinkage caused by the penalty terms.  

\subsection{Inverse problem solution}\label{sec:recon}

Minimizing \eqref{eq:penalty} poses a number of challenges. First of all, due to the nonlinearity of \eqref{eq:phase_velocity}, this function may not be convex and there are no guarantees that local optimization methods will converge to a global minimum.  Secondly, when the number of observations $T$ is large, this function may be costly to compute, and minimizing the number of function evaluations is paramount.

Fortunately, the theoretical challenge of nonconvexity does not prevent our method from obtaining consistently reliable model reconstruction, as we show in \sref{sec:results}. The computational difficulties can be addressed by using tools designed for neural networks and implemented in TensorFlow to efficiently compute gradients and perform minimization with a variant of mini-batch gradient descent \cite{github2019}.

\begin{figure}
    \centering
    \includegraphics[width=0.8\textwidth]{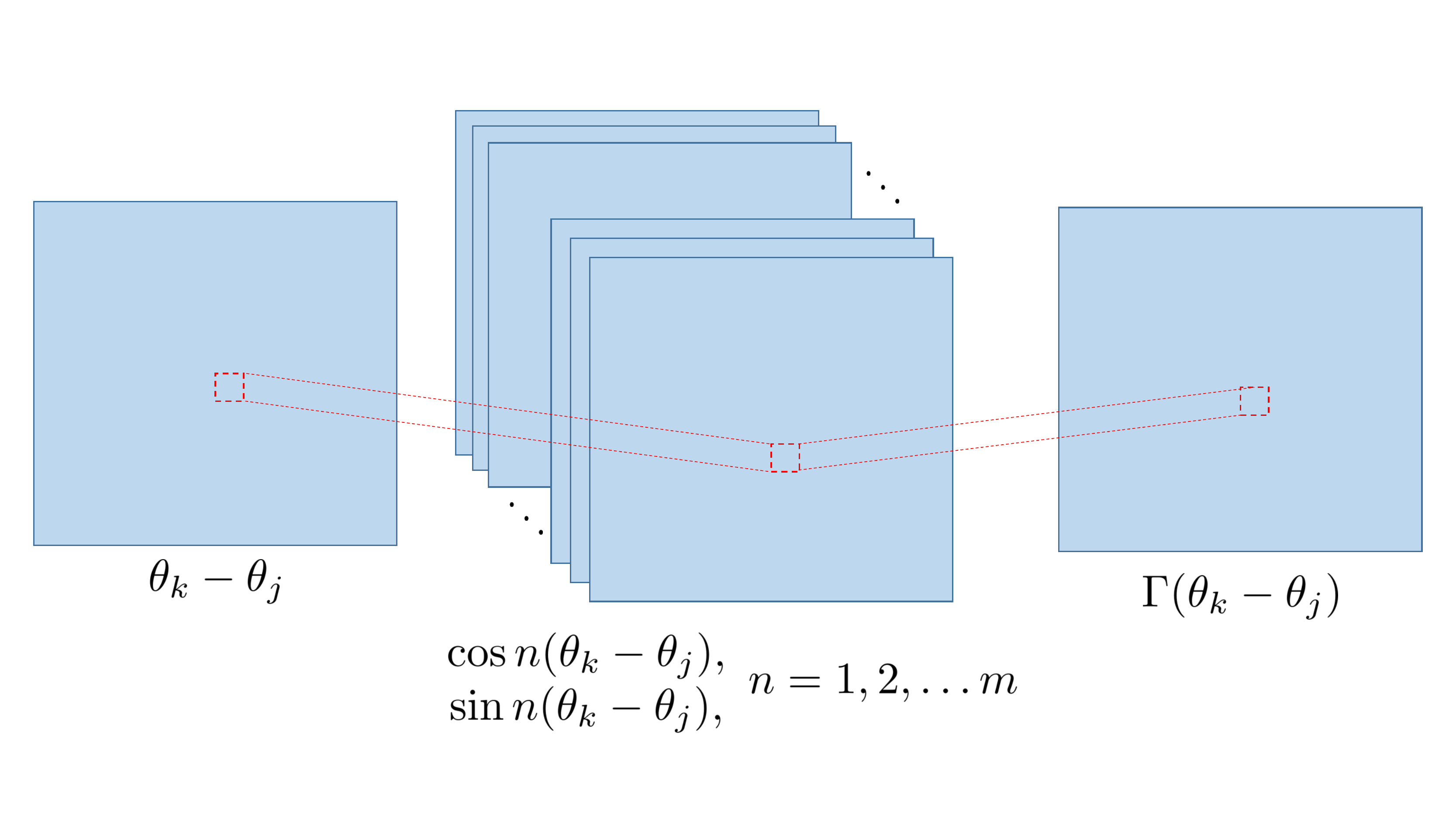}
    \caption{Coupling function as a neural network. This schematic illustrates how the coupling function $\Gamma$, given by \eqref{eq:fourier}, is evaluated at each phase difference $\theta_k-\theta_j$ via a convolutional neural network with size $1\times 1$ and stride 1. The hidden layer consists of $2m$ units with activation functions $\cos(n\theta)$ and $\sin(n\theta)$, $n=1,2,\ldots m$ with fixed inputs of weight 1.  The output layer uses variable weights representing the Fourier coefficients $a_n$ and $b_n$.}
    \label{fig:convolution}
\end{figure}

To elucidate the connection with neural networks, we comment that \eqref{eq:fourier} can be viewed as a 2-d convolutional neural network with a convolution of size $1\times 1$ and stride 1 applied to the $T\times N\times N$ tensor consisting of phase differences $\theta_k-\theta_j$. The hidden layer consists of the $2m$ harmonics of the form $\cos n\theta$ and $\sin n\theta$ applied to each entry in this tensor. To avoid redundant weights, we fix the weights in the first hidden layer to be 1 and the biases to be 0.  In the second hidden layer, each harmonic is assigned a fixed bias of 0 and a variable weight $a_n$ or $b_n$ corresponding to the Fourier coefficients. \Fref{fig:convolution} provides a schematic of this neural network. Once the coupling terms have been computed, the rest of \eqref{eq:phase_velocity} and the resulting loss function \eqref{eq:penalty} are straightforward to compute using vectorized operations.

We initialize the inferred parameters as follows: the adjacency matrix $A$ has initial entries drawn from $\mathcal{N}(0.5,1/N)$, the frequencies $\omega_k$ are initialized from $\mathcal{N}(0,1/N)$, the coupling strength $K$ is drawn from $\mathcal{N}(1,1/N)$, and the Fourier coefficients $\bm{a},\bm{b}$ of the coupling function are initialized at $0$.

TensorFlow maintains a computational graph for these operations which allows one to automatically compute gradients.  One can therefore use gradient descent methods to compute the optimal estimates ($\hat{a}_n$,$\hat{b}_n$,$\hat{\omega}_k$,$\hat{A}_{kj}$,$\hat{K}$) for the parameters. We used a mini-batch gradient descent method with batch size of $100$, \emph{i.e.}, we randomly assign the data to batches of 100 time-steps.  For each batch we calculate the gradient of the loss function and update the estimated parameters by taking a small step in a direction determined from the gradient. Once this has been repeated for all batches (one epoch), we pass through the data again for a total of $200$ or more epochs. We determined the number of epochs experimentally by iterating until the mean squared error no longer decreased. Typically, 300 epochs were sufficient. However for certain sweeps examining the role of $t_{max}$, $N_{res}$, and $\sigma$, the number of epochs required for convergence was as high as 10000.

By using small random batches of time steps, the gradient can be estimated quickly. This has the added benefit of introducing stochasticity into gradients which makes the algorithm less susceptible to getting caught in local minima. We use AdamOptimizer, which is a gradient descent method with momentum and an adaptive learning rate \cite{kingma2014}, with default parameter values for optimization in TensorFlow. 

\begin{table}
\caption{Default values and summary of method parameters for the results in \sref{sec:results}.}
\label{tab:methodparams}
\begin{indented}
\item[]\begin{tabular}{@{}l*{15}{l}}
\br
Parameter & Description & Value \\
\mr
n\_epochs & iterations through the training data & 300 \\
batch size & time-steps of data per gradient descent batch & 100 \\
$m$ & number of inferred Fourier coefficients & 5 \\
$\lambda_\Gamma$ & weight of penalty on non-sparse coupling & 0.0001 \\
$\lambda_A$ & weight of penalty on network connectivity & $10^{-6}$ \\
$\lambda_{bd}$ & weight of penalty on $A_{kj}\not\in [0,1]$ & $10^5$ \\
\br
\end{tabular}
\end{indented}
\end{table}

There is minor variability in the success of the algorithm due to the random initialization of the inferred parameters and to the batching of the observed data.  In order to ensure an accurate network reconstruction, we can retrain the model with new initial values several times and choose the result which has the smallest mean squared error for the velocity predictions on the validation set, which was not considered during the training process. For any individual model network, this validation error is correlated with the accuracy of the inferred parameters.  In all examples below, we attempt to reconstruct each network five times before choosing the best reconstruction.

\subsection{Post-processing}\label{sec:postproc}
The method described in \sref{sec:recon} produces continuous estimates for the parameters ($\hat{a}_n$,$\hat{b}_n$,$\hat{\omega}_k$,$\hat{A}_{kj}$,$\hat{K}$) which could be used to predict the dynamics of the system under a variety of initial conditions.  However, we are interested in evaluating whether these parameter estimates are an accurate reconstruction of the original model.  Before these values can be compared to the model parameters used to generate the data, additional post-processing is needed.  We therefore redefine our parameter estimates via the transformations $\hat{K}\hat{A}_{kj}\rightarrow \hat{A}_{kj}$, $c_0+c_1\hat{\Gamma}\rightarrow \hat{\Gamma}$ where $c_0$ and $c_1$ are selected to minimize $\displaystyle\int_0^{2\pi} |\Gamma(\theta)-\hat{\Gamma}(\theta)|d\theta$, and $\hat{\omega}_k-K{c_0}/{N}\sum_{j=1}^NA_{kj}\rightarrow \hat{\omega}_k.$ These transformation are necessary to address the aforementioned identifiability issues with $K$ and to allow $\Gamma$ to have nonzero mean. 

In the original model, the adjacency matrix entries are either zero or one. However, we treat the entries as continuous variables during optimization and then choose a threshold $\epsilon$ so that $A_{kj}<\epsilon$ is chosen to be 0 and $A_{kj}\geq \epsilon$ is chosen to be 1. In practice, one could fix $\epsilon$ or select $\epsilon$ so that the reconstructed model minimizes the mean squared error for the data set.  However, we explore a range of threshold values using ROC curves and report the reconstruction error rates using the value of $\epsilon$ that yields the largest $F_1$ score for the adjacency matrix reconstruction (see \ref{app:perf_metrics}). These optimal values are robust and good performance is typically obtained over a wide range of thresholds (see \sref{sec:results} for details).

\section{Experimental design\label{sec:experiments}}
In order to validate the robustness of our approach, we test the reconstruction method on data simulated with a variety of networks and parameter values.  We explore this high-dimensional parameter space by fixing all of the model parameters except for one and then sweeping the remaining parameter over a wide range of values. See tables \ref{tab:networkparams} and \ref{tab:solutionparams} for a list of default values as well as the ranges considered. For each set of parameter values, we compute 30 networks with random initial phases and intrinsic frequencies.  We then attempt to reconstruct the underlying model for each network and compute various performance metrics for each reconstruction.  

\textit{Coupling function} --- We evaluate the inferred coupling function $\hat{\Gamma}(\theta)$ by comparing it to the true coupling function $\Gamma(\theta)$ via the \textit{normalized difference in area} defined as follows:

\begin{equation}
    \mathrm{Area \; ratio} = \frac{\displaystyle\int_0^{2 \pi} |\Gamma(\theta) - \hat{\Gamma}(\theta)|d\theta}{\displaystyle\int_0^{2 \pi} |\Gamma(\theta)|d\theta} \,.
\label{eq:area}
\end{equation}
This represents the area between the true and estimated coupling function curves, weighted by the area under the curve of the true function. This quantity serves as a measure of the error in the reconstruction. A perfect reconstruction would have a normalized difference in area of zero. In the Kuramoto case $\Gamma(\theta) = \sin(\theta)$, the initial estimate $\hat{\Gamma}(\theta)=0$ corresponds to a value of 1. Therefore values significantly lower than 1 indicate progress towards a correct reconstruction.

\textit{Intrinsic Frequencies} --- We compare the inferred intrinsic frequencies $\hat{\omega}_k$ to the true intrinsic frequencies $\omega_k$  using the \textit{mean absolute deviation} defined as follows:
\begin{equation}
\mathrm{Mean \; absolute \; deviation} = \frac{1}{N}\sum_{k=1}^N |\omega_k - \hat{\omega}_k|.
\label{eq:freq}
\end{equation}
Values near zero indicate accurate reconstructions. We considered alternative metrics such as relative deviations and the correlation between true and inferred frequencies, but these were less informative because they tend to amplify errors when the the intrinsic frequencies are close to zero.

 \begin{figure}[h!]
 \begin{center}
 \sidesubfloat[]{\hspace{0.1cm}
 \includegraphics[height=0.27\textwidth,clip=true,trim= 3cm 0cm 3cm 0cm]{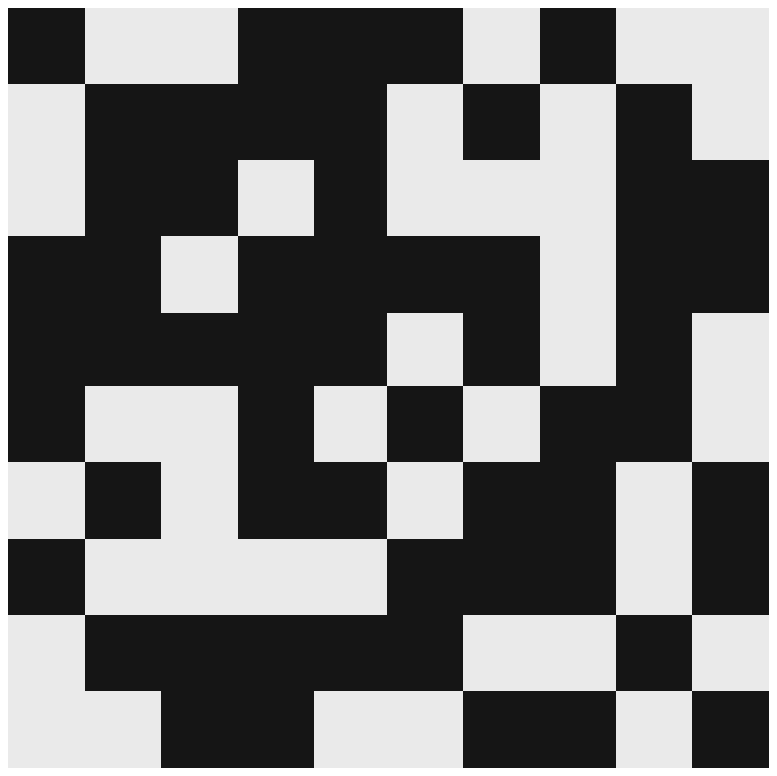}~}
 \sidesubfloat[]{\hspace{0.2cm}
 \includegraphics[height=0.27\textwidth,clip=true,trim= 3cm 0cm 1.5cm 0cm]{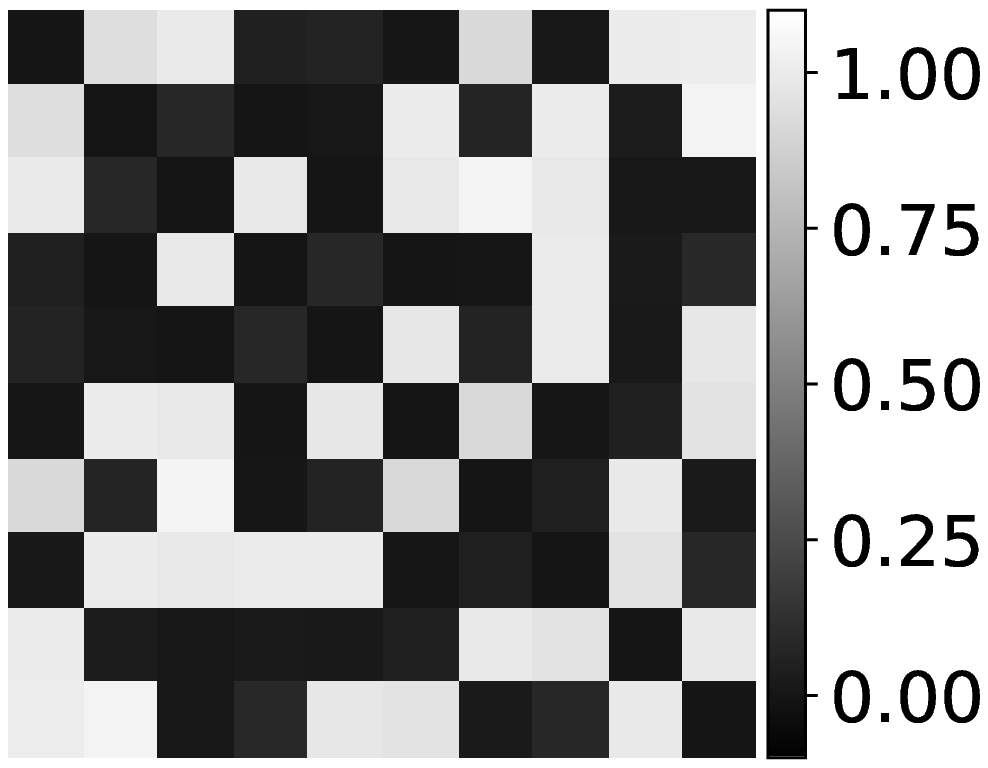}~~}
 \sidesubfloat[]{\hspace{0.1cm}
  \includegraphics[height=0.27\textwidth,clip=true,trim= 3cm 0cm 1.5cm 0cm]{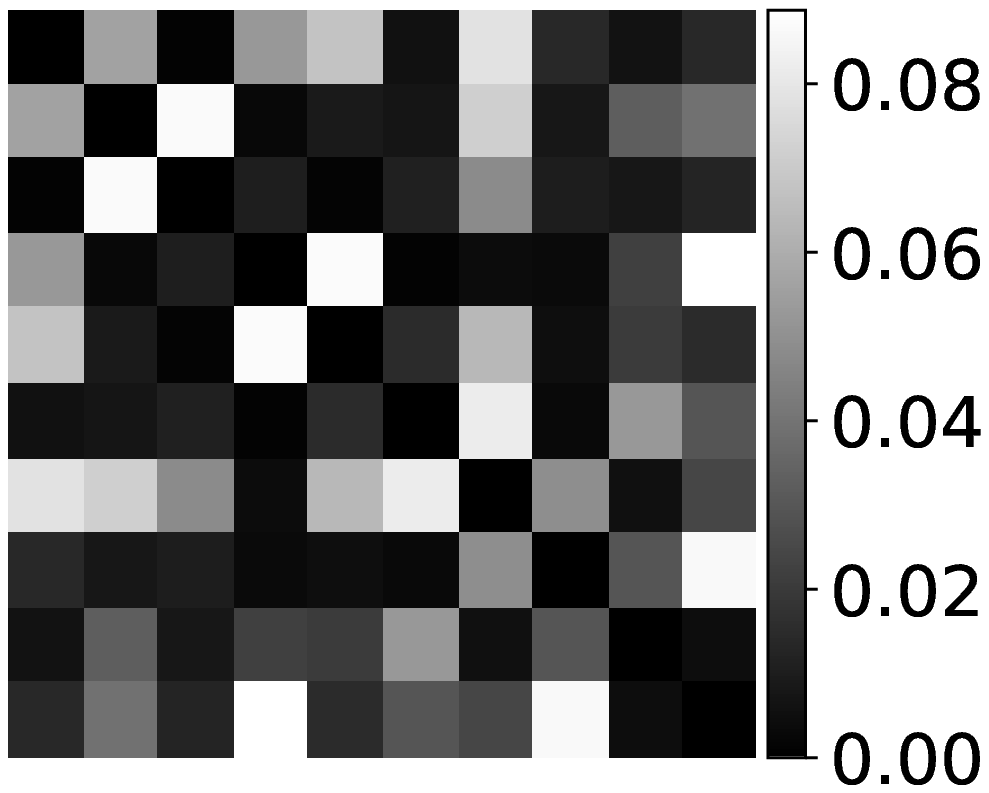}~}
  \end{center}
  \begin{center}
  \sidesubfloat[]{\hspace{-0.3cm}
  \includegraphics[height=0.29\textwidth,clip=true,trim=0.5cm 0cm 0cm 0cm]{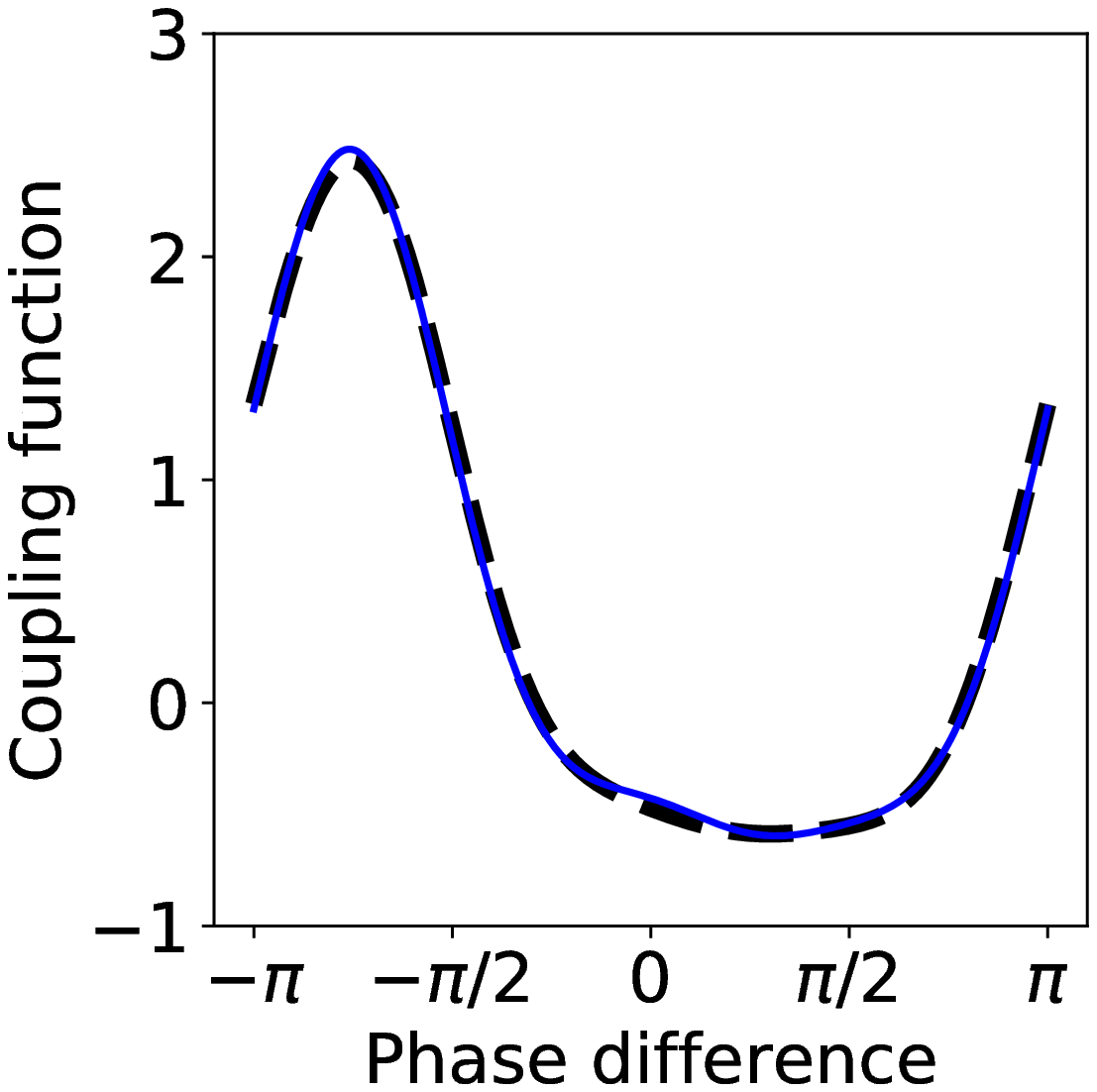}~~} \sidesubfloat[]{\hspace{-0.3cm}
  \includegraphics[height=0.29\textwidth,clip=true,trim=0.5cm 0cm 0.8cm 0cm]{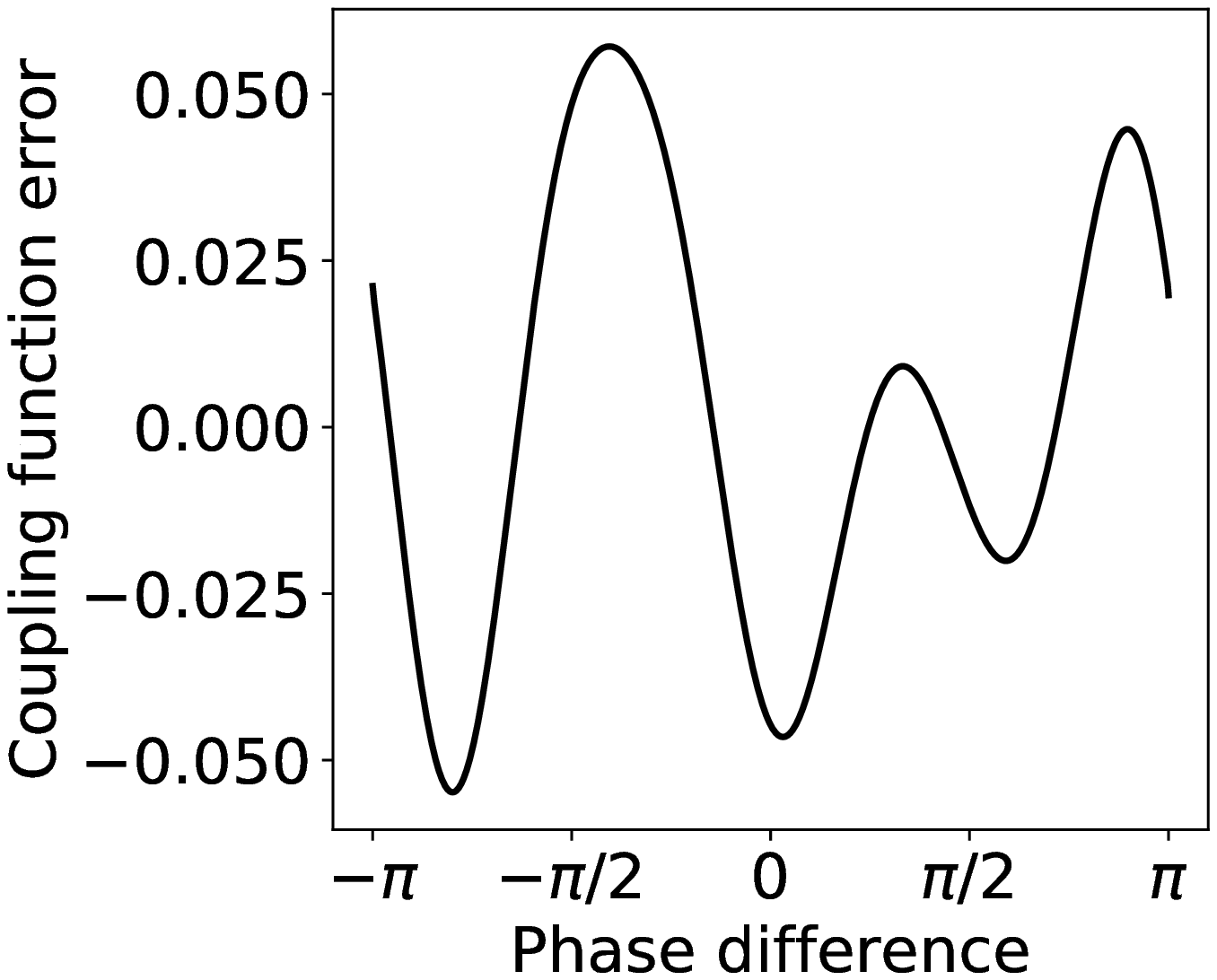}~}
  \sidesubfloat[]{\hspace{-0.3cm}
    \includegraphics[height=0.29\textwidth,clip=true,trim=0.5cm 0cm 0cm 0cm]{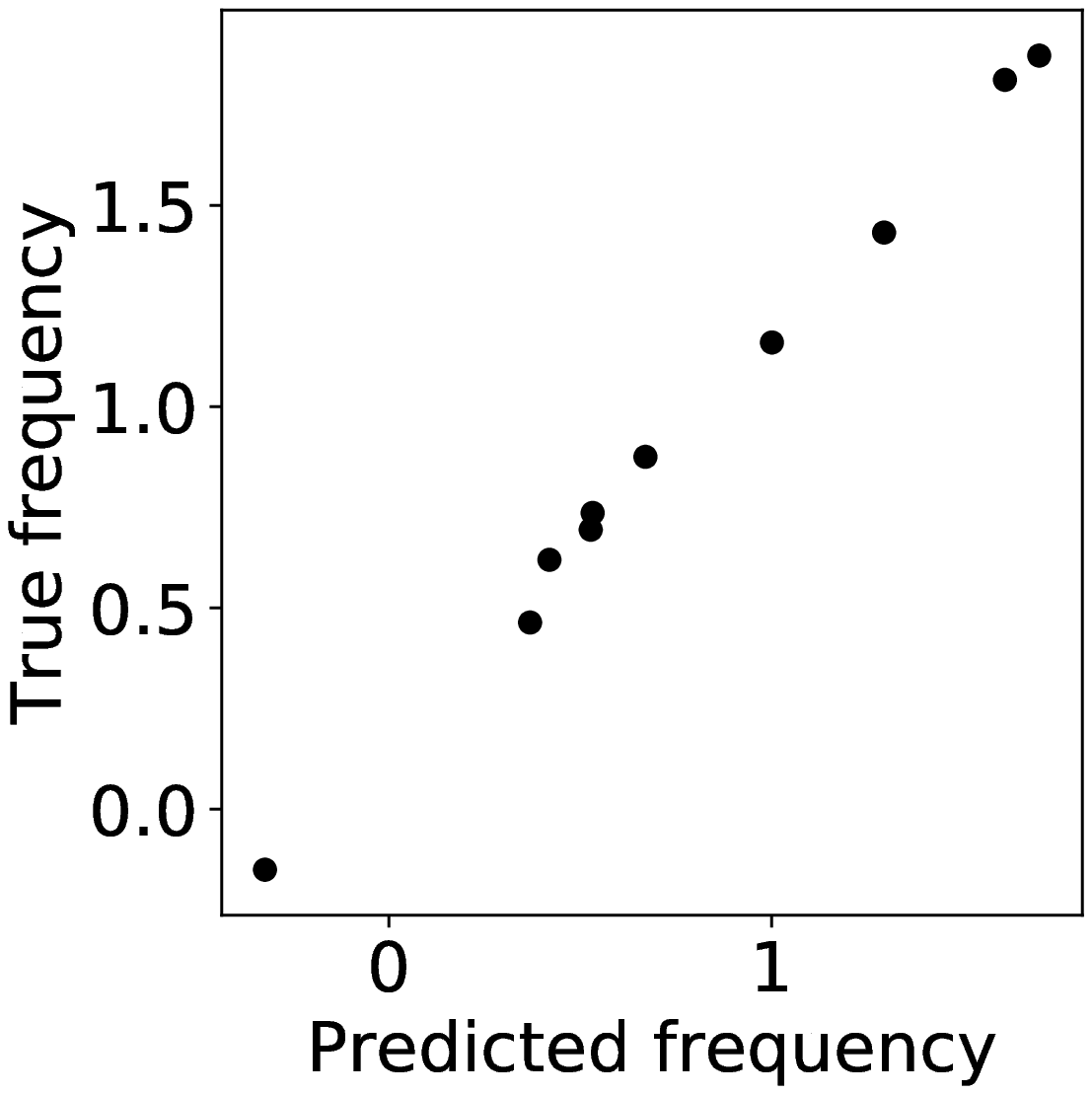}~~~~}
\end{center}
 \caption{Predicted adjacency matrix, coupling function, and frequencies. (a) True adjacency matrix; (b) predicted adjacency matrix;  (c) absolute difference between true and predicted adjacency matrices; (d) coupling functions, true (blue, solid) and predicted (black, dashed); (e) coupling function difference; (f) predicted vs.\ true frequencies.
 Here we obtain an adjacency matrix classification error rate of 0 for thresholds between 0.1 and 0.9, a normalized difference in area of 0.032 for the coupling function, and a mean absolute deviation of 0.008 for the estimated natural frequencies.}
\label{fig:example data}
 \end{figure}

\textit{Adjacency matrix} --- We investigate several methods for evaluating the success of adjacency matrix reconstruction. The accuracy can be inspected visually by plotting the true and reconstructed matrices along with the absolute differences in a grid where values range from $0$ (black) to $1$ (white); see \fref{fig:example data}(a-c). Since we are primarily interested in discrete adjacency values, we interpret reconstruction as a classification problem and compute three standard evaluation metrics: the \textit{$F_1$ score}, the classification \textit{error rate} for the connections, and the \textit{area under the ROC curve}. See \ref{app:perf_metrics} for precise definitions of these metrics.  

\section{Results}\label{sec:results}

\begin{figure}[t!]
 \centering
   \sidesubfloat[]{\includegraphics[width=0.45\textwidth]{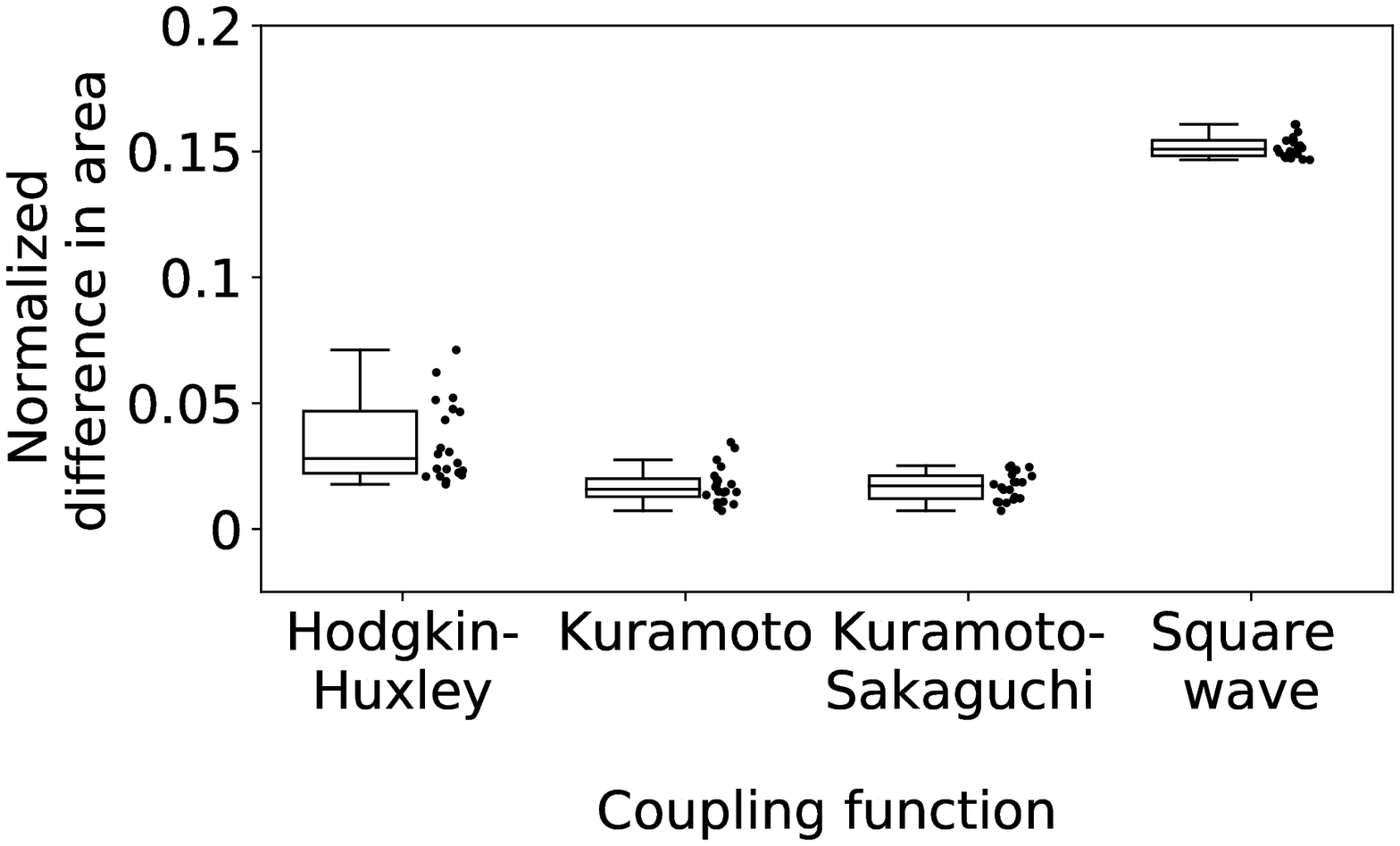}\label{fig:coup_sweep}}
  \sidesubfloat[]{\includegraphics[width=0.45\textwidth]{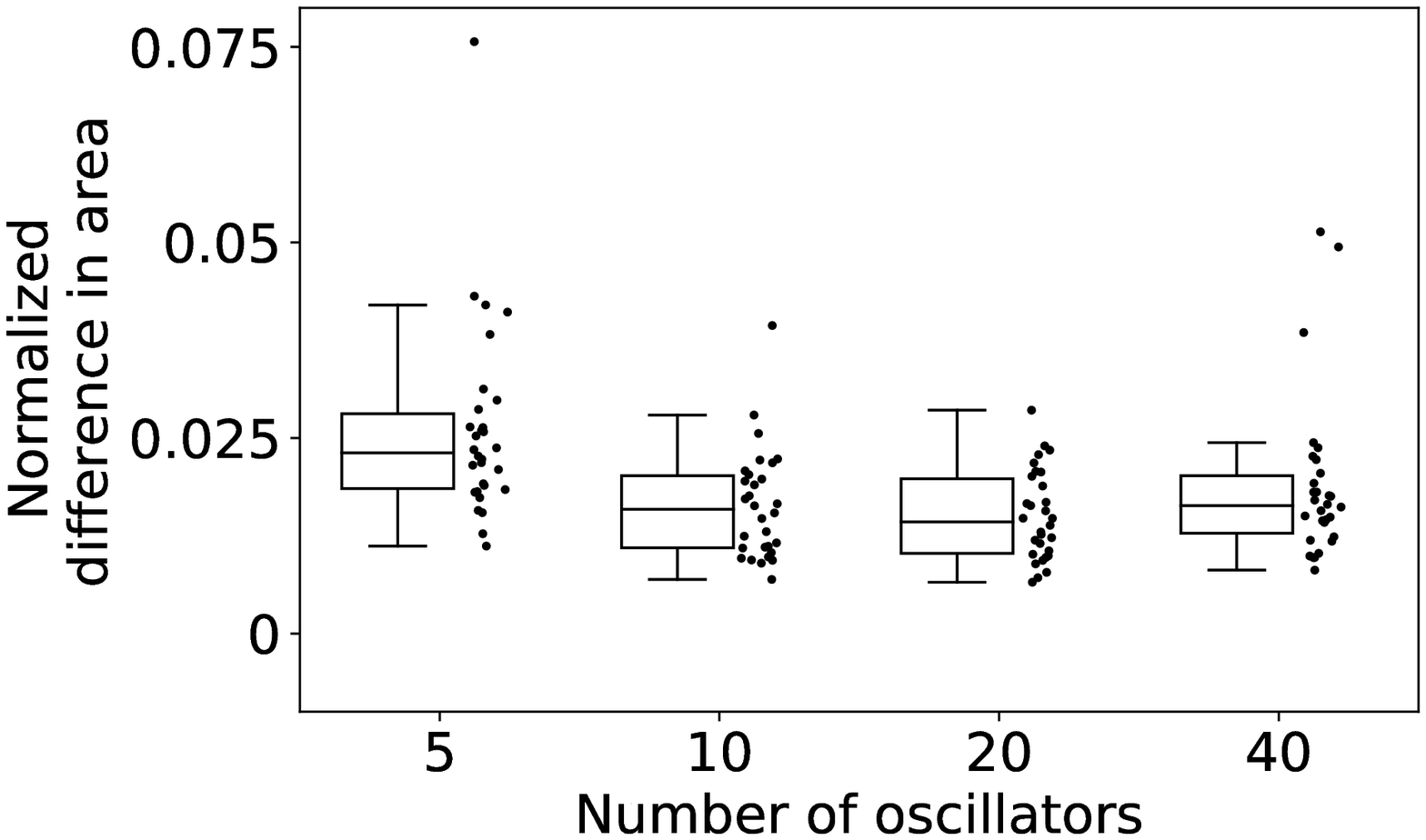}\label{fig:num_sweep}}

 \caption{Performance of coupling function reconstruction for different coupling functions and varying numbers of oscillators.  (a) Normalized difference in area for different coupling functions. For the Hodgkin-Huxley, Kuramoto, and Kuromoto-Sakaguchi coupling functions, the normalized difference in area has the order $O(10^{-2})$. For a square-wave coupling function, the normalized difference in area is larger, but still reasonably small ($O(10^{-1})$). Whiskers are 1.5 times the interquartile range (IQR).
 (b) Normalized difference in area \eqref{eq:area} for different number of oscillators $N$. As the number of oscillator increases, the normalized difference in area remains small.
  }
 \label{fig:coupling}
\end{figure}

Here we discuss the key results from the parameter sweeps outlined in \sref{sec:experiments}. 

Our method can successfully reconstruct models with a variety of coupling functions. \Fref{fig:coup_sweep} and \tref{tab:coupling_fcn} show the results for four different coupling functions (see \ref{app:coup_funcs} for details), three of which come from popular coupled oscillator models: the Kuramoto model, the Kuramoto-Sakaguchi model, and a phase reduction of the Hodgkin-Huxley model, and a fourth, a square wave which represents a generic discontinuous function with high frequency Fourier harmonics. We note that for the square wave coupling function, the accuracy of the reconstruction suffers since the reconstruction includes only five Fourier harmonics and the true coupling function contains an infinite number of harmonics with amplitudes that decay slowly due to the discontinuities. Despite these limitations, the adjacency matrix was still estimated with a high degree of accuracy.

Counterintuitively, as the number of oscillators increases (\fref{fig:num_sweep} and \tref{tab:num_osc}), we find that reconstruction accuracy improves.  This can be explained by the observation that although increasing the number of oscillators increases the number of unknown parameters, it also provides a greater number of pairwise phase differences which can aid in reconstruction of the coupling function. The resulting modest improvements in the estimate of the coupling function can then allow for better inference of the adjacency matrix and intrinsic frequencies as well. This is illustrated by the normalized difference in area plot in \fref{fig:num_sweep}.

\begin{figure}[t!]
 \sidesubfloat[]{
 \includegraphics[width=0.45\textwidth]{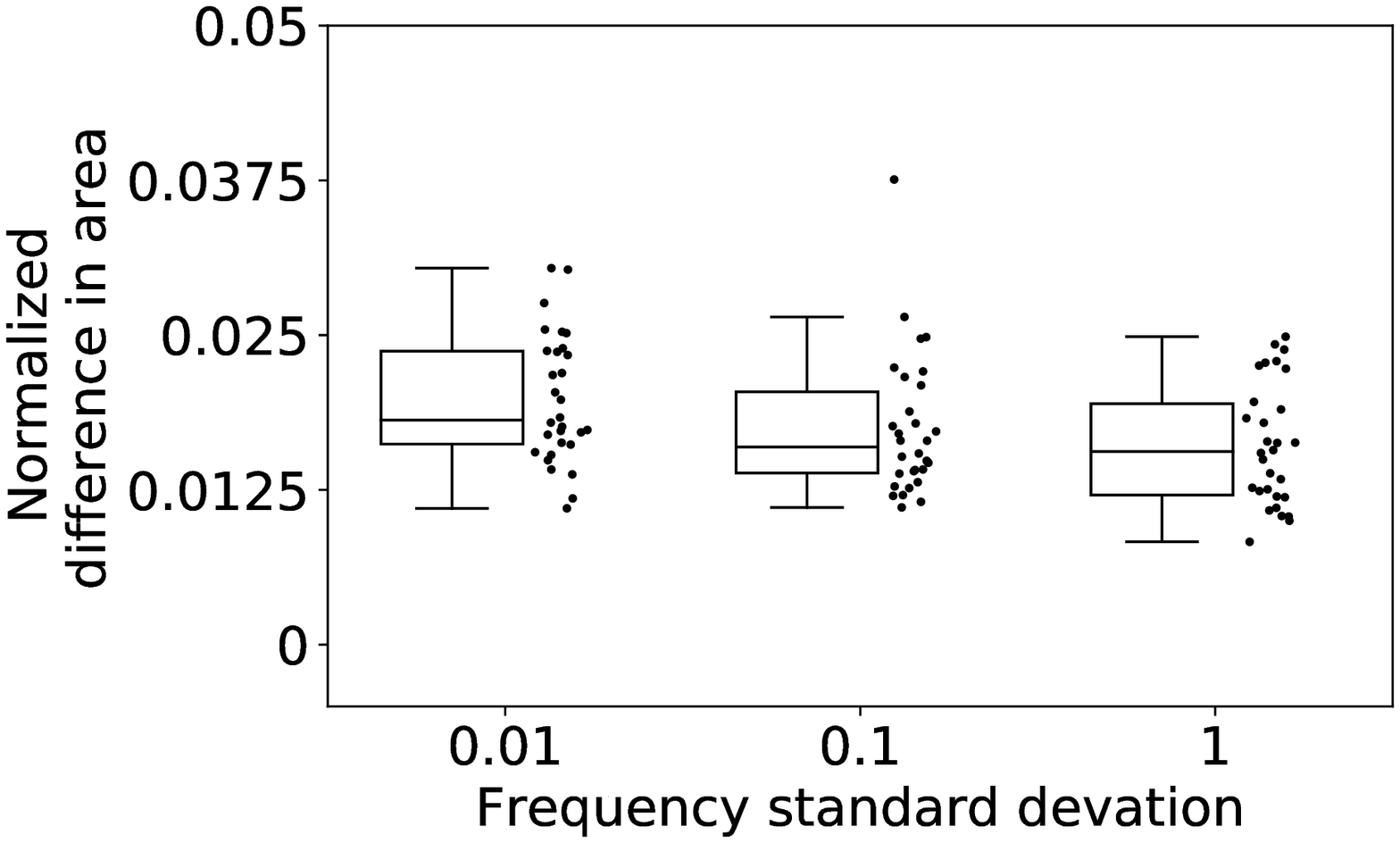}}
 \sidesubfloat[]{
 \includegraphics[width=0.45\textwidth]{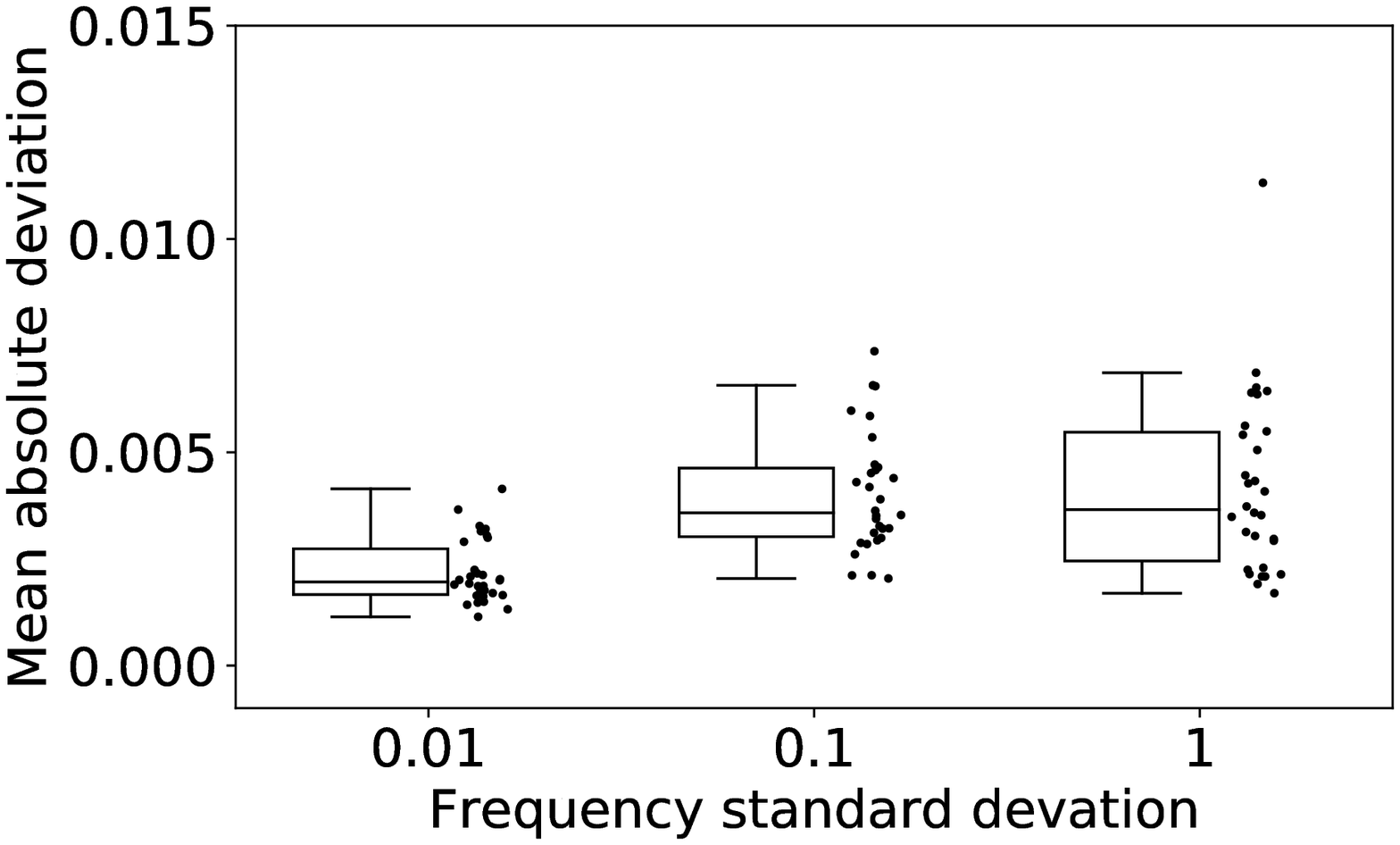}}
 \caption{Performance metrics of the coupling function and the intrinsic frequencies for various frequency standard deviations. (a) Normalized difference in area. As the frequency standard deviation changes from 0.01 to 1, the normalized the difference in area remains small. (b) Mean absolute deviation of the frequencies \eqref{eq:freq}. A small nonzero deviation in the intrinsic frequencies tends to reduce the mean absolute deviation. }
\label{fig:frequency}
 \end{figure}

In \fref{fig:frequency} and \tref{tab:sigma_freq}, we consider simulations where the standard deviation $\sigma$ of the oscillator frequencies ranges from $0.01$ to $1$. The normalized difference in area of the coupling function and the mean absolute deviation of the inferred frequencies both show that we can infer the network well for $\sigma\leq 1$. For larger standard deviations, the quality of the reconstruction suffers slightly, since the large frequencies dominate the coupling terms within the phase velocity relationship. 

  \begin{figure}[t!]
 \sidesubfloat[]{
 \includegraphics[width=0.45\textwidth]{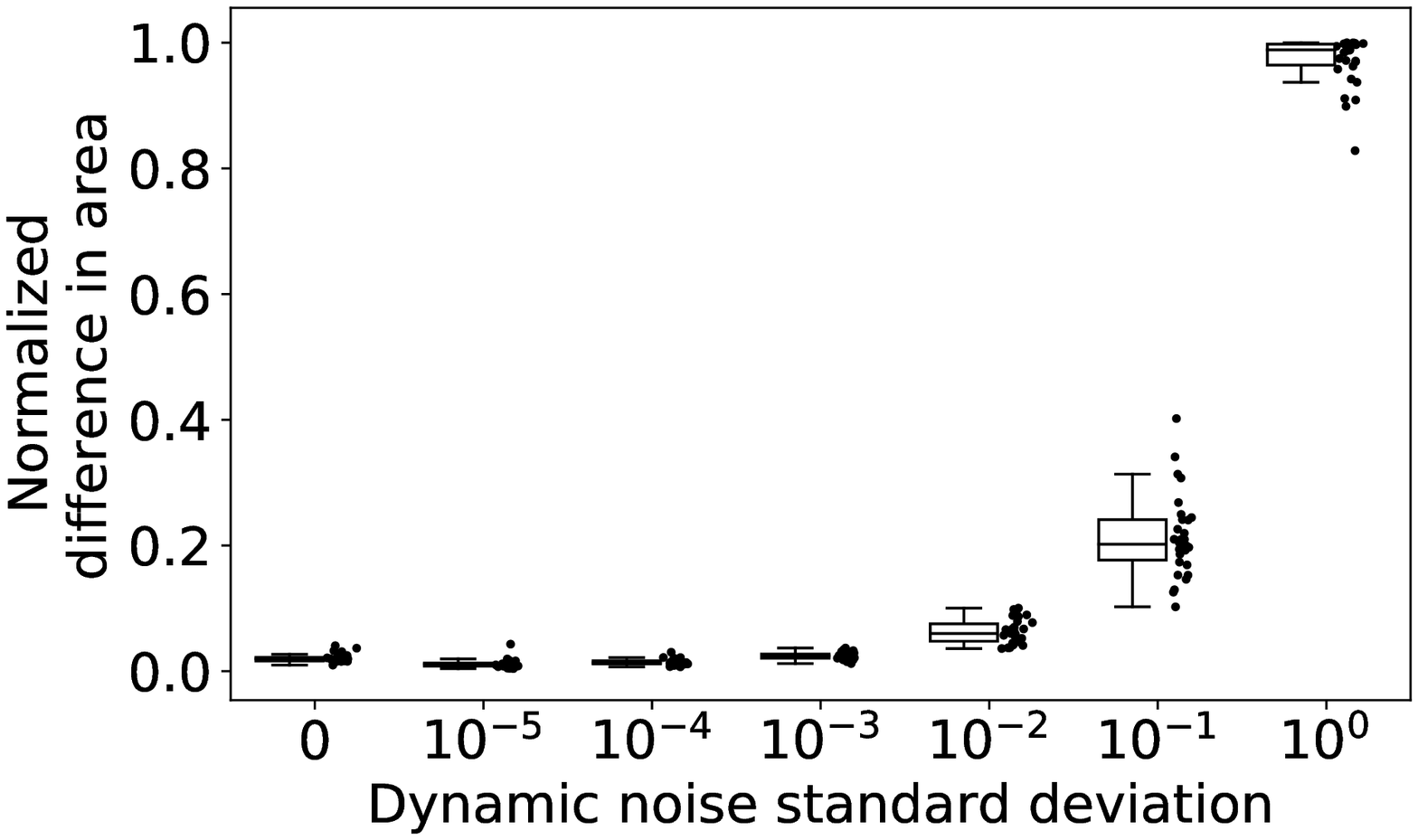}}
 \sidesubfloat[]{
 \includegraphics[width=0.45\textwidth]{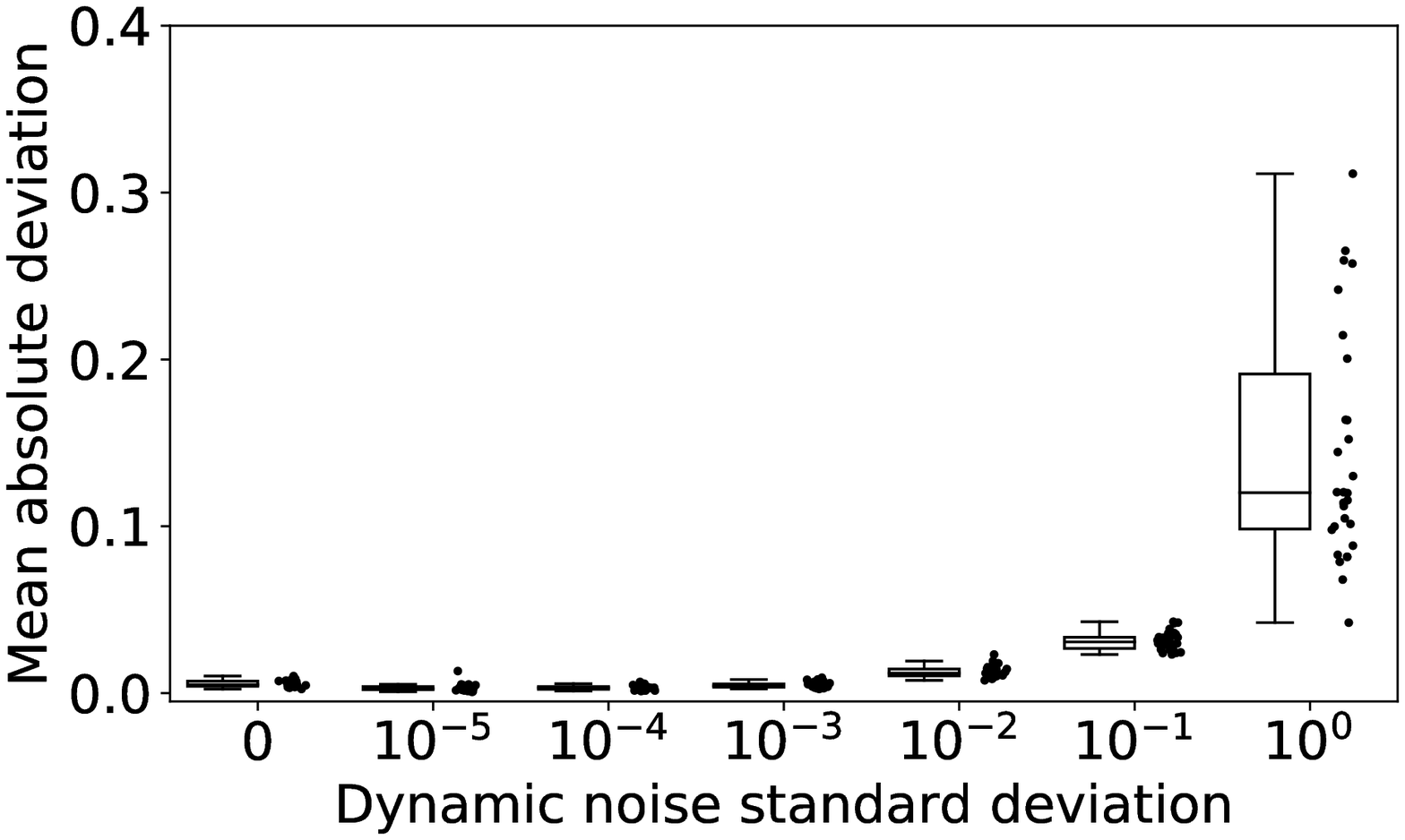}}
 \caption{Response of reconstruction to dynamic noise with various standard deviations. (a) Normalized difference in area of the coupling function. When the dynamic noise standard deviation is less than $10^{-3}$, the normalized difference in area is smaller than for a model without noise. As the dynamic noise standard deviation increases to $10^{-1}$, we observe an increase in the normalized difference in area. A significant increase is observed as the noise standard deviation increases to 1. (b) Mean absolute deviation of the frequencies. The mean absolute deviation remains small as the dynamic noise standard deviation increases to $10^{-1}$.}
\label{fig:noise}
 \end{figure}

We also explore the impact of noise on the model reconstruction since both observation noise and dynamic noise would be present in experiments. In \tref{tab:noise_level} and \tref{tab:dynamic_noise_level}, we demonstrate that our model reconstruction method is robust to moderate amounts of both types of noise. As in \cite{ShaTim2011}, we introduce dynamic noise using stochastic differential equations with Gaussian noise. Observation noise was also Gaussian and was added after integrating the governing differential equations. For both types of noise, we used mean $0$ and standard deviations between $0$ and $1$. It is worth noting that dynamic noise with a standard deviation less than $10^{-3}$ helps with the reconstruction of both coupling function and frequencies. This can be explained by the fact that a small amount of noise keeps the system from reaching equilibrium, effectively increasing the duration of the transients that provide useful information about the structure of the network. 

We carry out additional parameter sweeps with varying network connectivity $p$ (\tref{tab:p_erdos}),  maximum simulation time $t_{\max}$ (\tref{tab:tmax}) and the number of simulation restarts (\tref{tab:full_refresh}) with all other parameters set using the default values given in tables \ref{tab:networkparams}, \ref{tab:solutionparams}, and \ref{tab:methodparams}. In each case, we are consistently able to reconstruct the coupling function, the underlying network and the intrinsic frequencies model over a wide range of model parameters. The tables in \ref{app:tables} illustrate averaged numerical results for 30 trials of each of these parameter sweeps in terms of evaluation metrics such as normalized difference in area, mean absolute deviation, error rate, area under ROC curve, and a range of thresholds that yield $F_1$ scores within $90\%$ of the largest value.

\subsection{Results for perturbations of synchronous dynamics}\label{subsec:perturbations}
As both \cite{ShaTim2011} and \cite{Pik2018} point out, when a network remains synchronized, i.e. when $\dot{\theta}_1=\dot{\theta}_2=\ldots=\dot{\theta}_N$, one cannot infer the model parameters due to the fact that the observed phases no longer provide linearly independent equations. Furthermore, even with precise knowledge of the adjacency matrix, one could not hope to reconstruct the coupling function $\Gamma(\theta_k-\theta_j)$ without data over a wide range of phase differences $\theta_k-\theta_j$. 

\begin{figure}[p!]%[htb!]
  % 12 - fixed 1, random size 0.01 "control"
  \sidesubfloat[]{\hspace{-0.2cm}
   \includegraphics[width=0.31\textwidth]{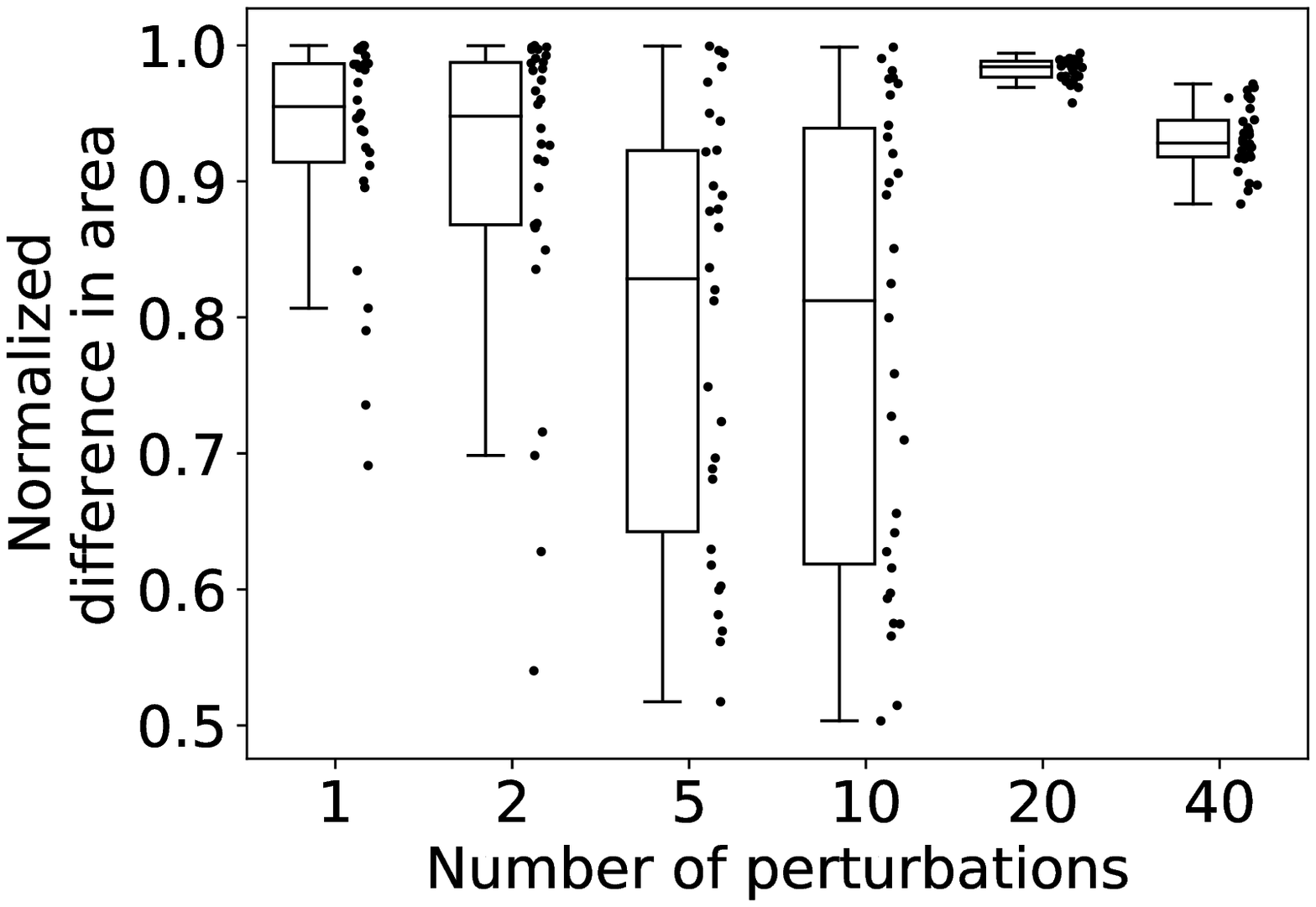}}
  \sidesubfloat[]{\hspace{-0.2cm}
 \includegraphics[width=0.31\textwidth]{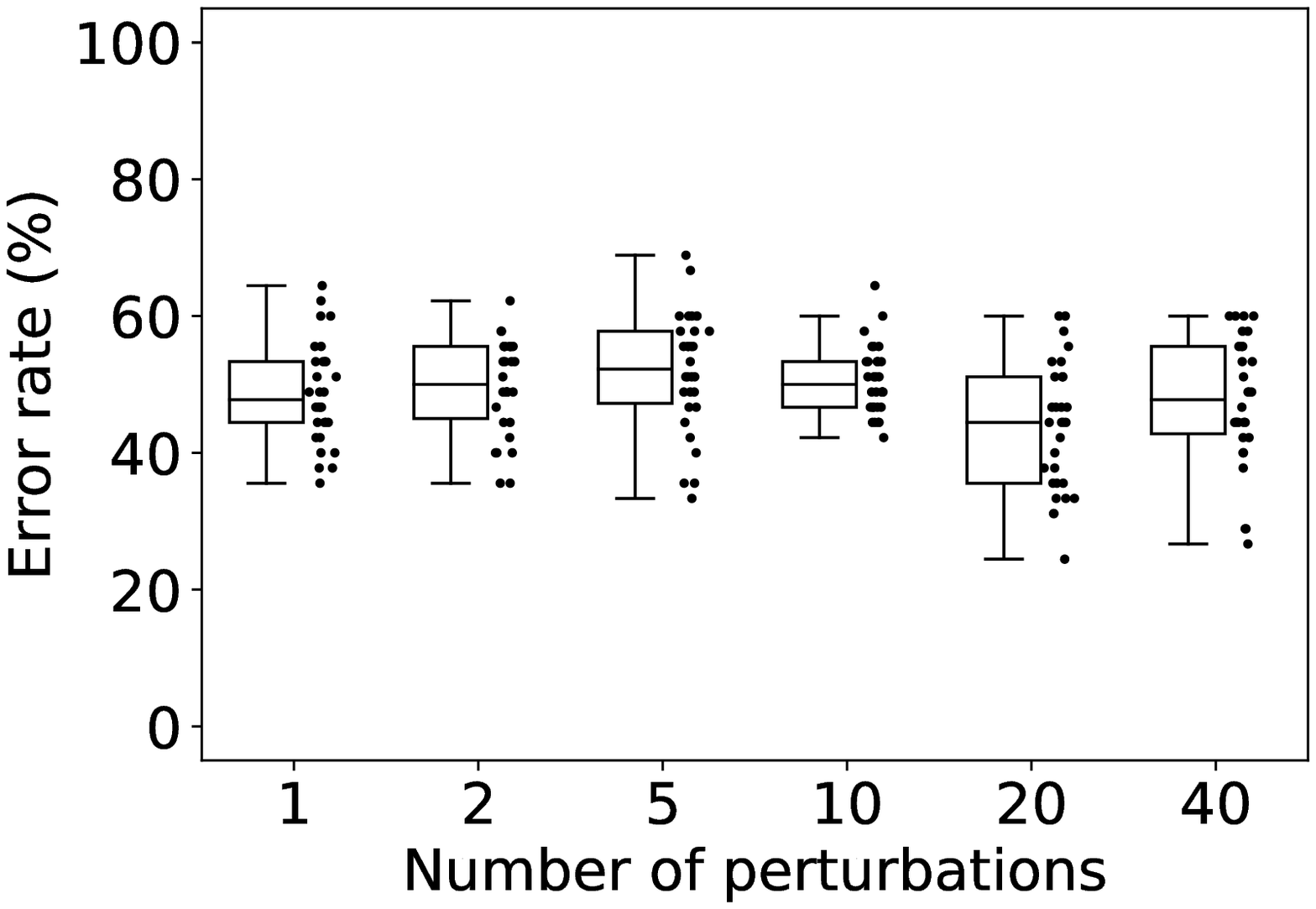}}
 \sidesubfloat[]{\hspace{-0.2cm}
  \includegraphics[width=0.31\textwidth]{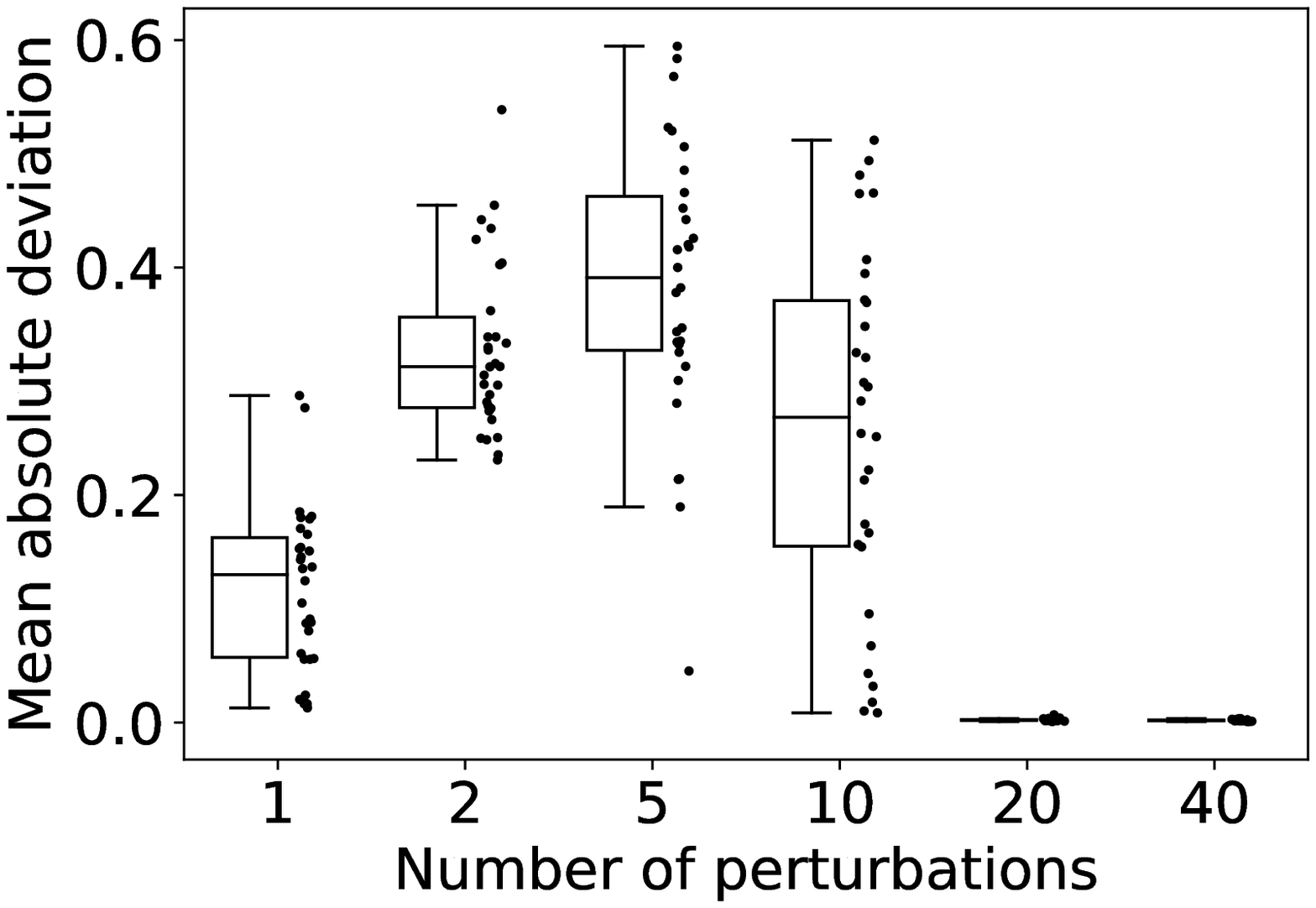}}
  \hfill
  % 23 - fixed 3, random size 10
  \sidesubfloat[]{\hspace{-0.2cm}
   \includegraphics[width=0.31\textwidth]{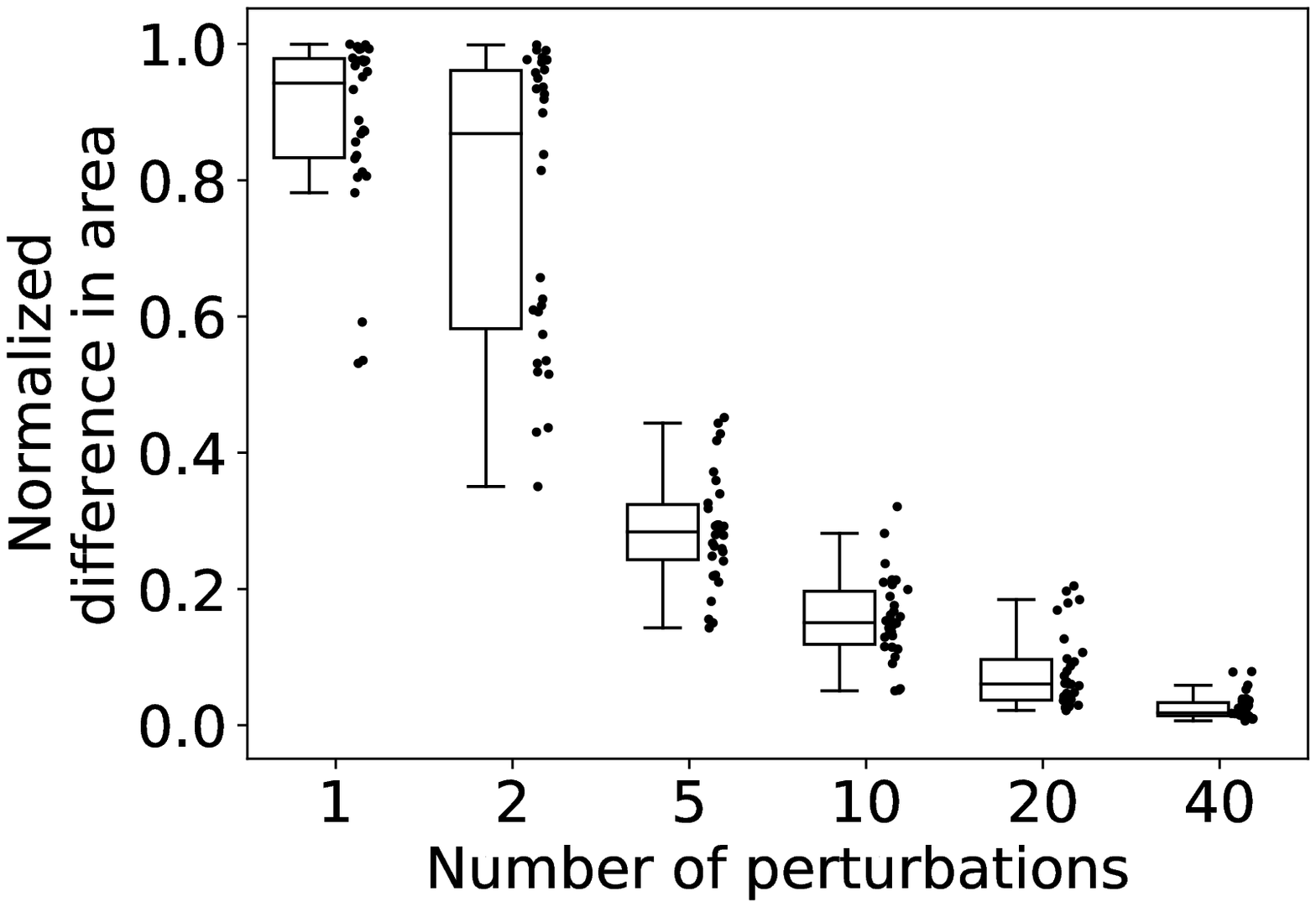}}
   \sidesubfloat[]{\hspace{-0.2cm}
 \includegraphics[width=0.31\textwidth]{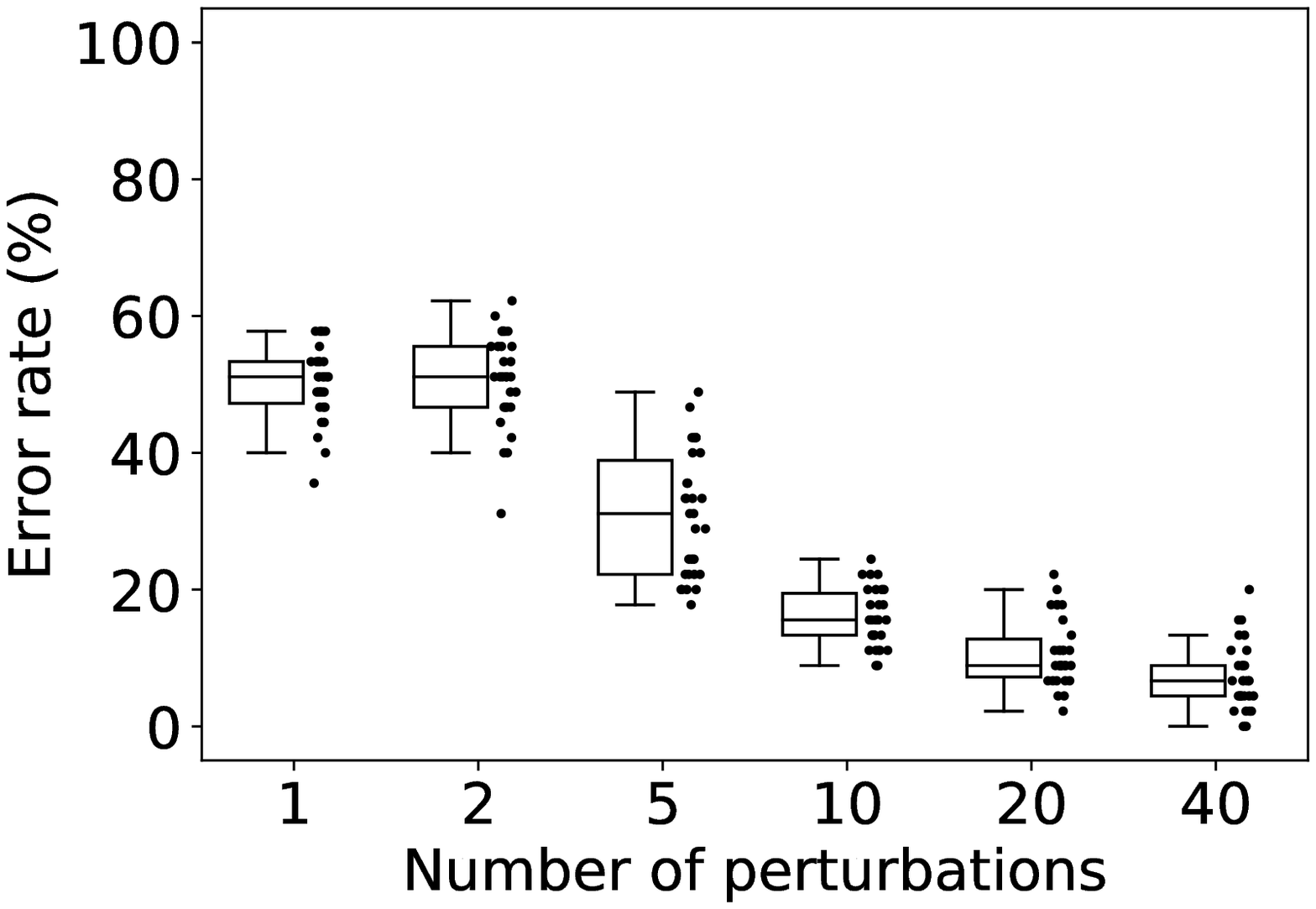}}
 \sidesubfloat[]{\hspace{-0.2cm}
  \includegraphics[width=0.31\textwidth]{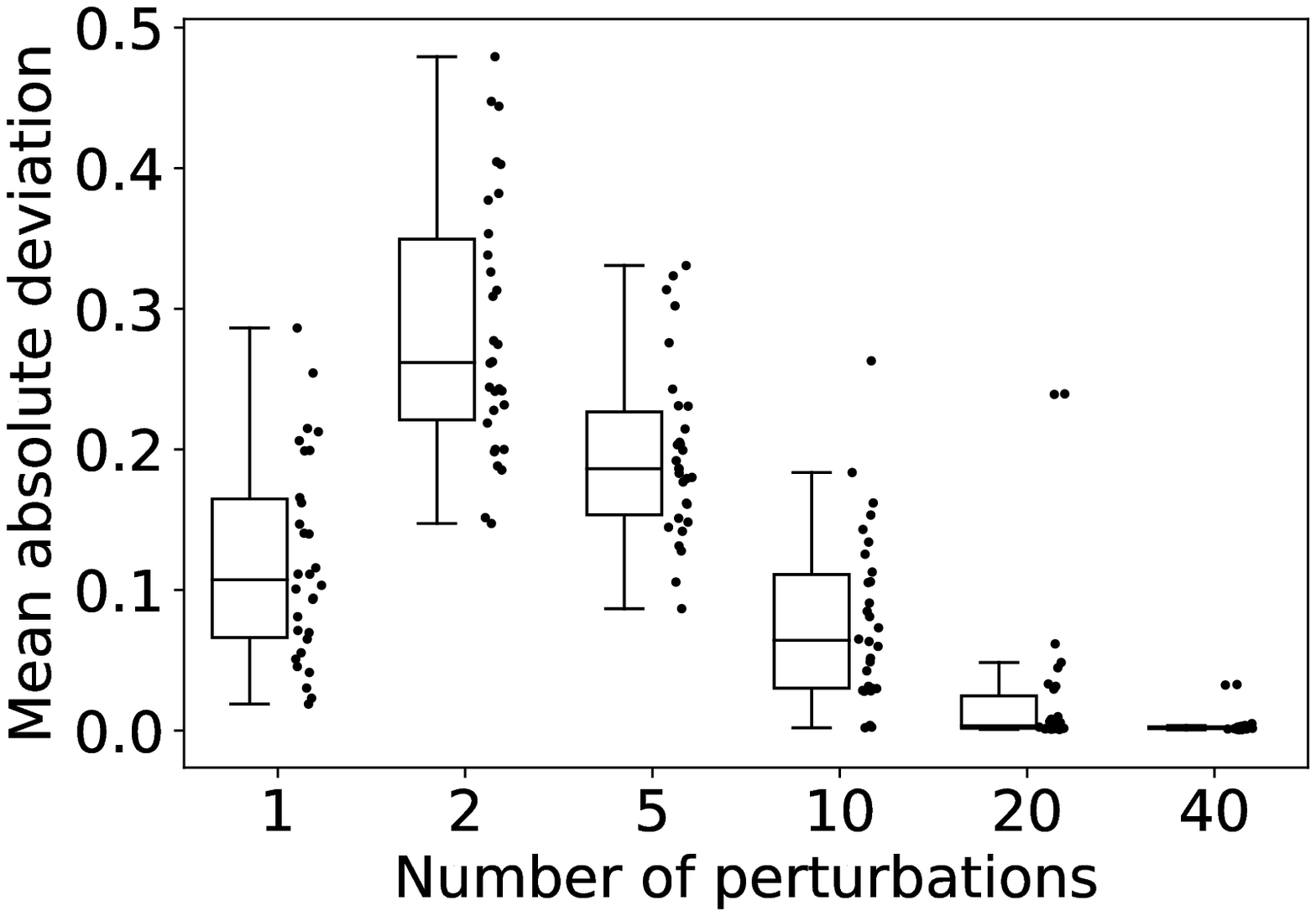}}
  \hfill
  % 33 - random 1, random size 10
  \sidesubfloat[]{\hspace{-0.2cm}
   \includegraphics[width=0.31\textwidth]{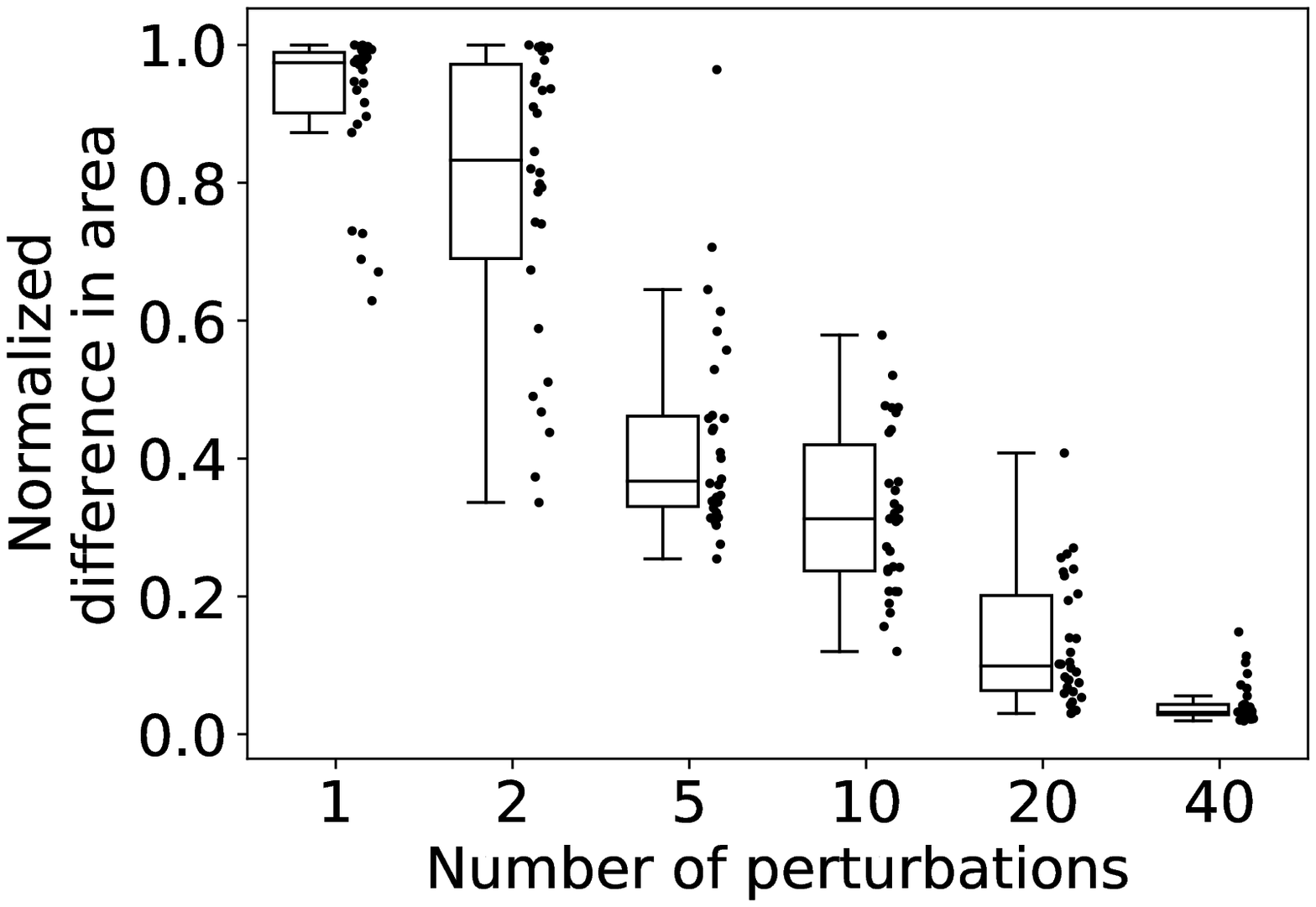}}
   \sidesubfloat[]{\hspace{-0.2cm}
 \includegraphics[width=0.31\textwidth]{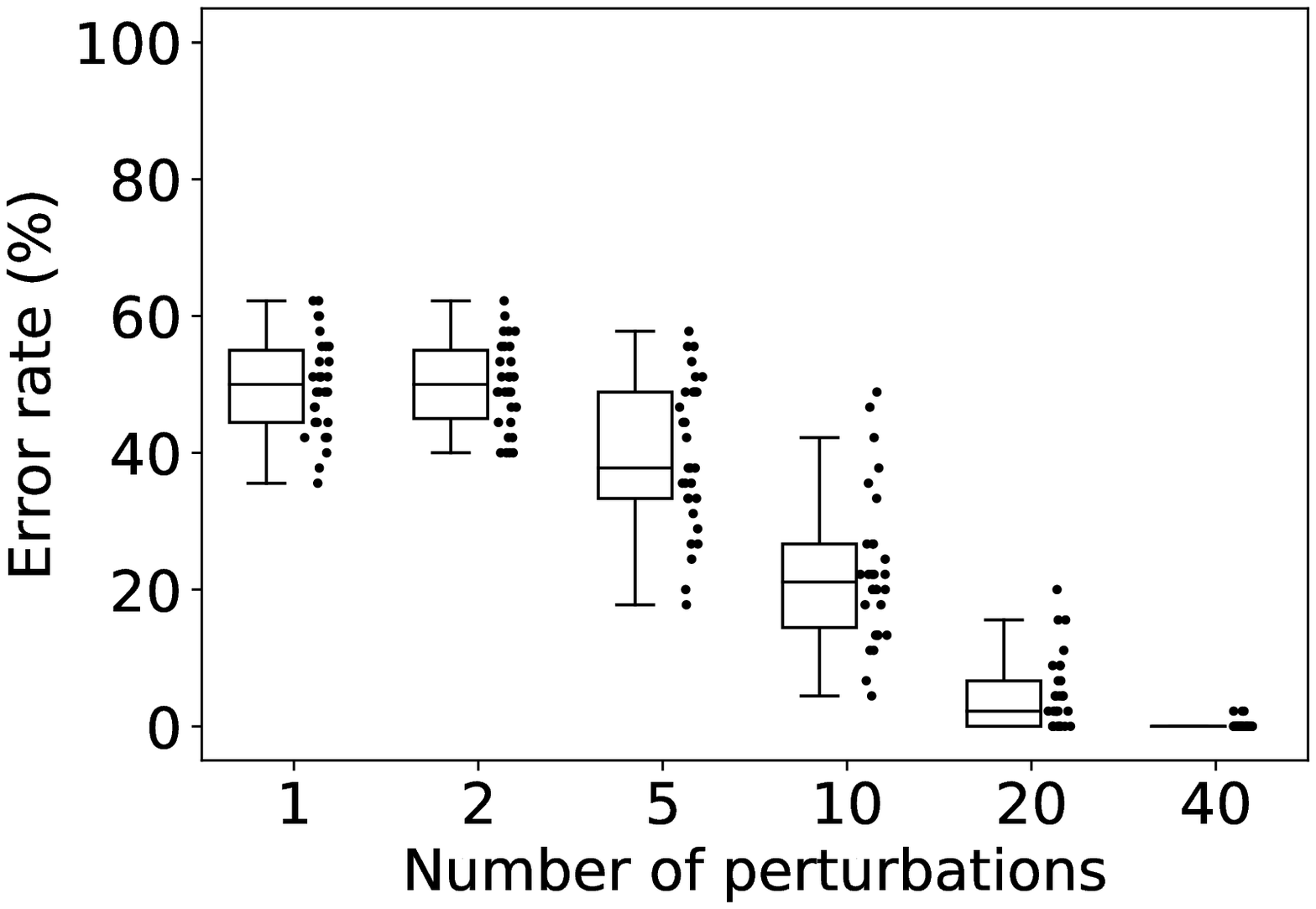}}
 \sidesubfloat[]{\hspace{-0.2cm}
  \includegraphics[width=0.31\textwidth]{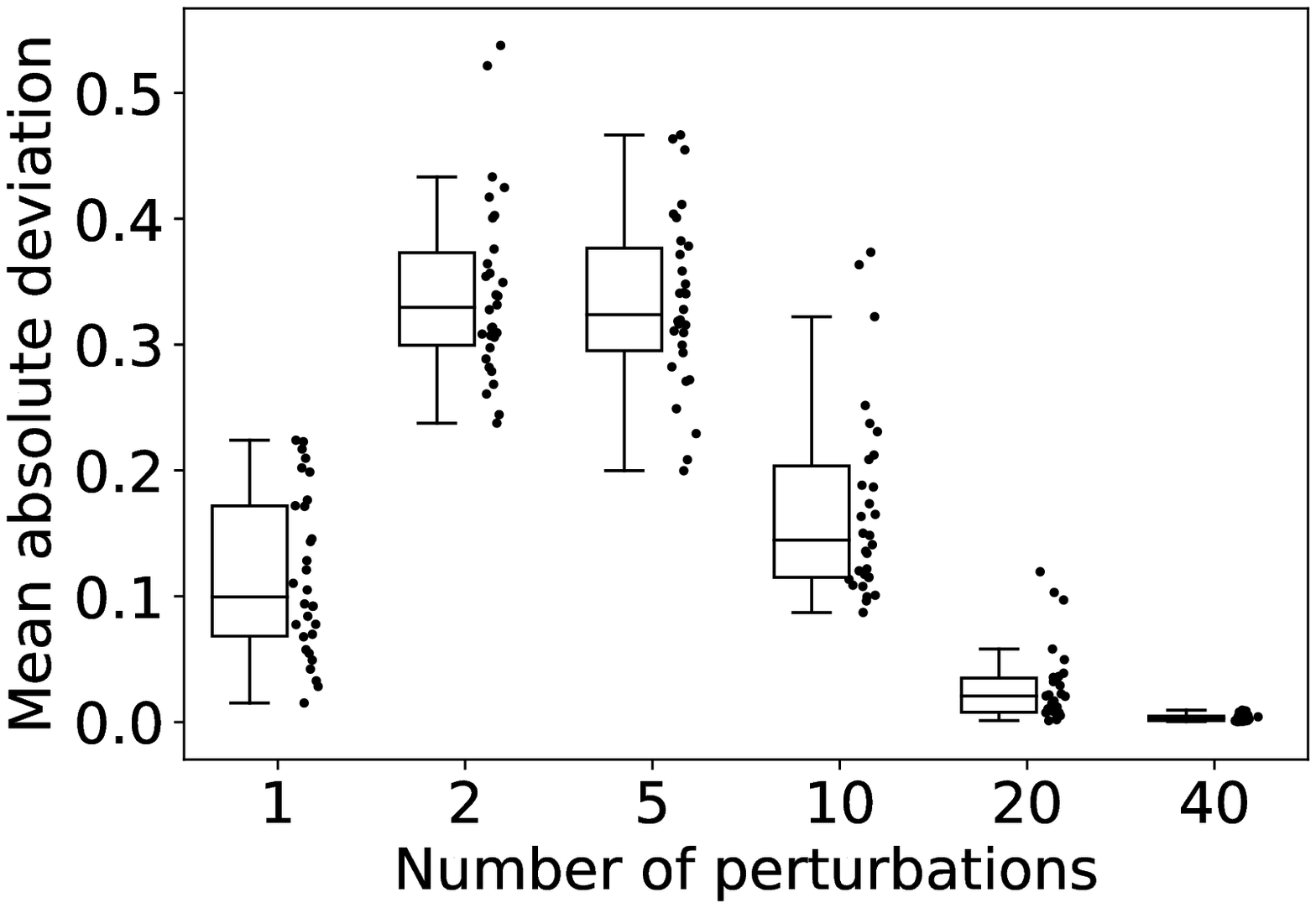}}
  \hfill
  % 11 - random 3, reset
  \sidesubfloat[]{\hspace{-0.2cm}
 \includegraphics[width=0.31\textwidth]{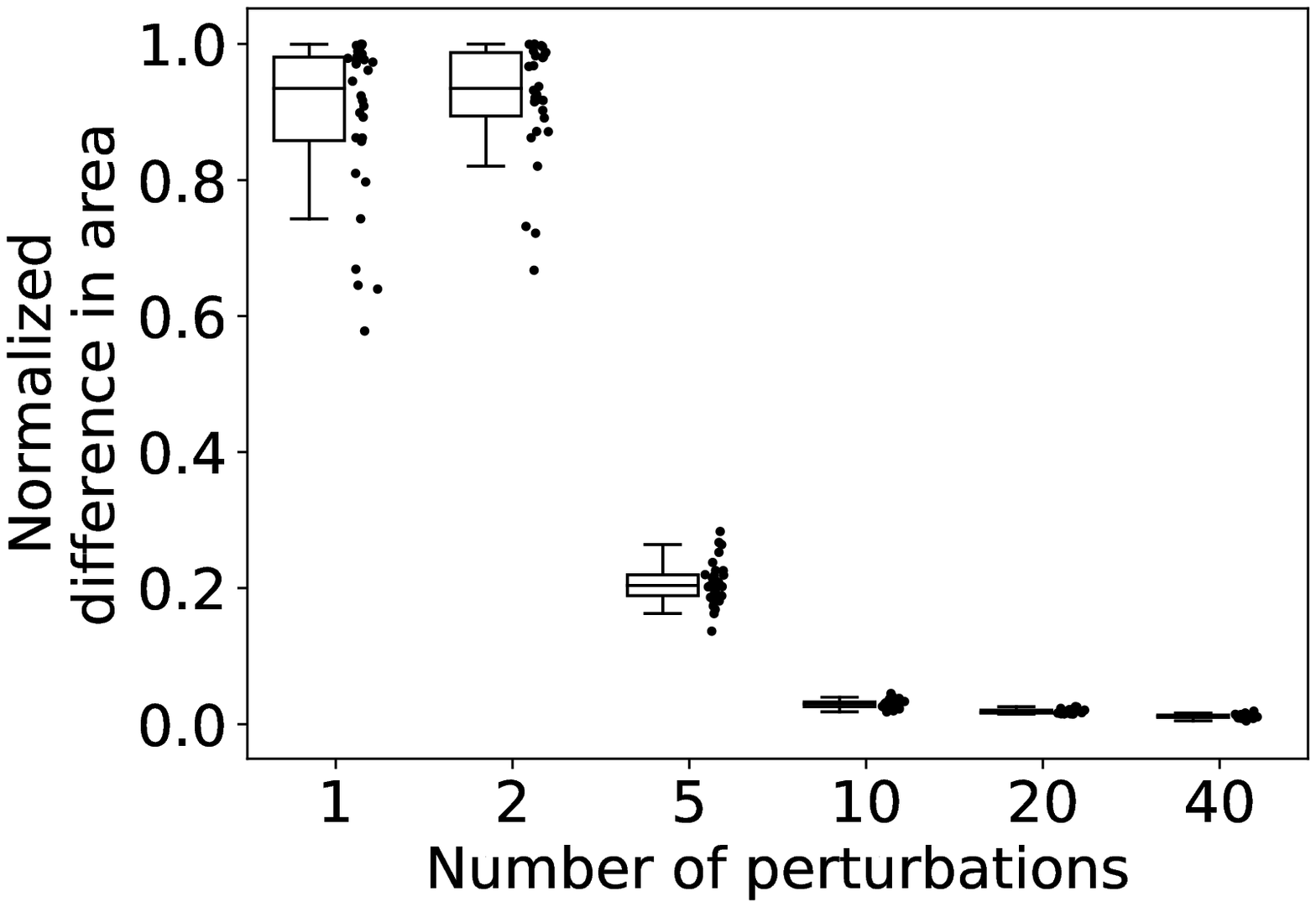}}
 \sidesubfloat[]{\hspace{-0.2cm}
 \includegraphics[width=0.31\textwidth]{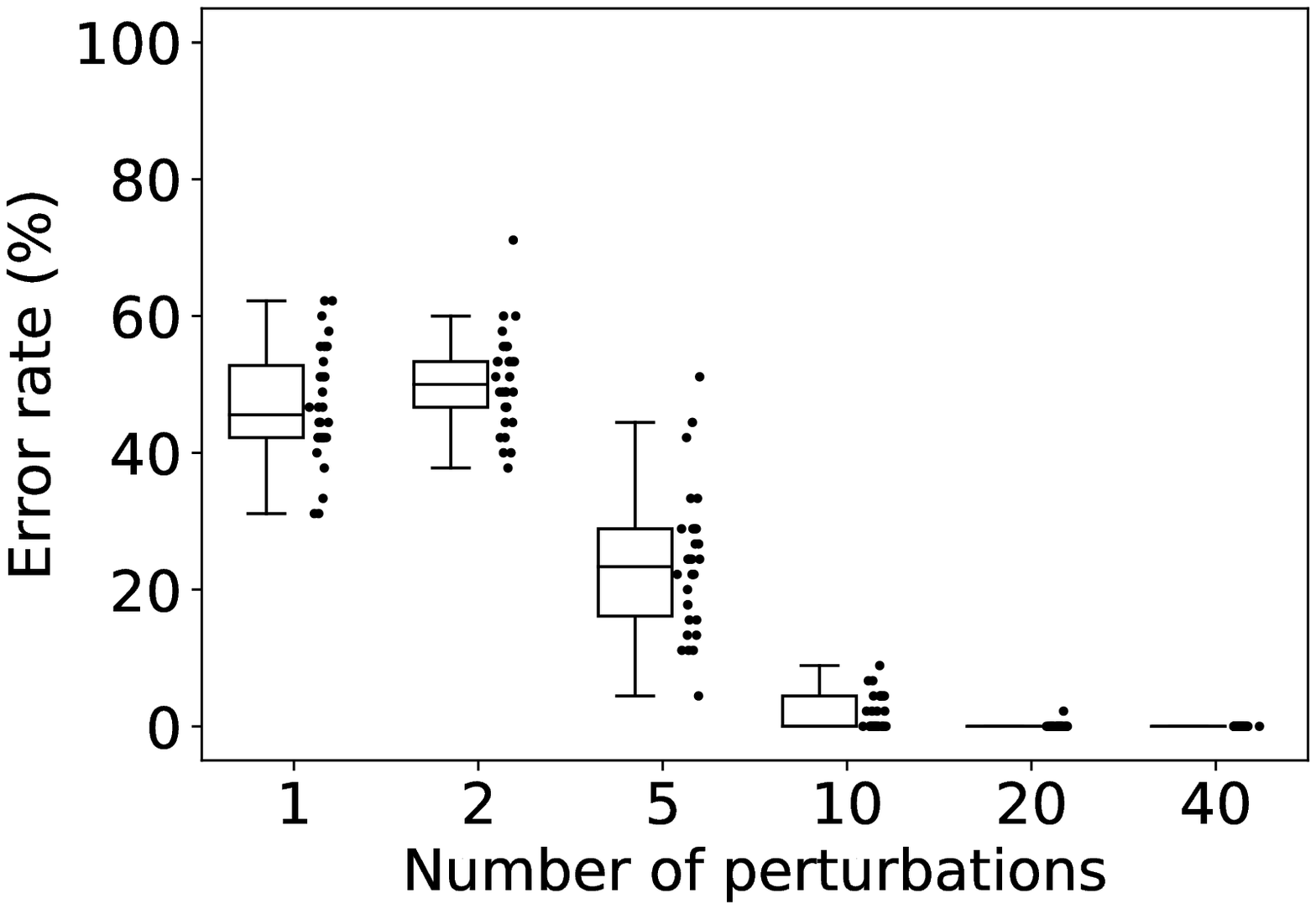}}
 \sidesubfloat[]{\hspace{-0.2cm}
  \includegraphics[width=0.31\textwidth]{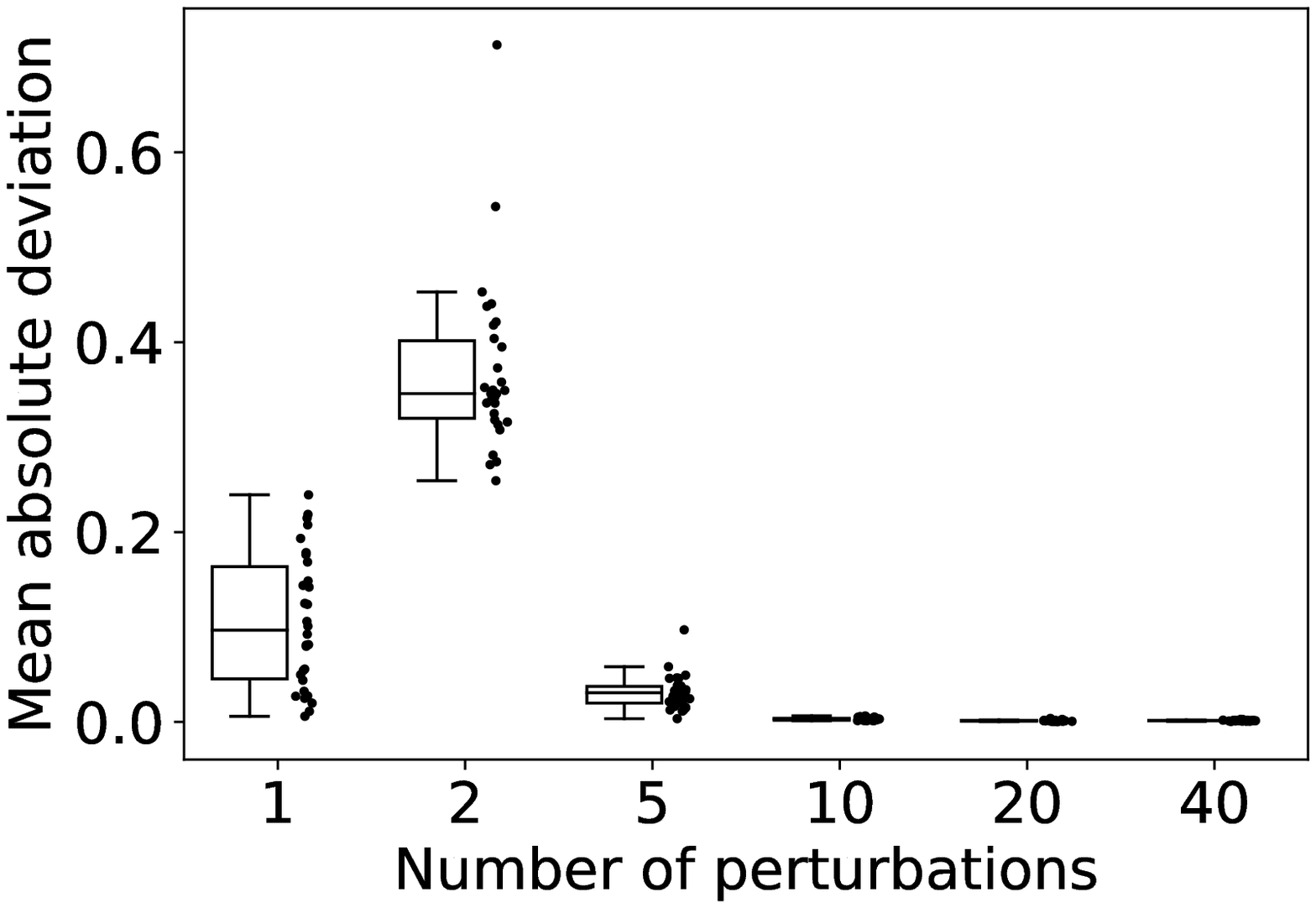}}
 \caption{Reconstruction of a model with synchronous dynamics.  We simulate the dynamics for $t_{max}=200$ units of time and introduce perturbations at times $t={kt_{max}}/{N_{pert}}$ for $k=1,2,\ldots, N_{pert}$.  Panels (a-i) use random normally distributed phase perturbations $\eta_{pert}\sim \mathcal{N}(0,\sigma_{pert}^2)$ with (a-c) $\sigma_{pert}=0.01$ and (d-i) $\sigma_{pert}=10$  and panels (j-l) use phase resetting.  In panels (a-c) we perturb a fixed subset of 1 oscillator (out of 10) repeatedly.  In panels (d-f), we perturb a fixed subset of three oscillators repeatedly. In panels (g-i), we perturb a single oscillator selected randomly for each perturbation. In panels (j-k), we reset a subset of three oscillators selected randomly with each perturbation. The first column displays the normalized difference in area for the coupling function, the second displays the error rate for the adjacency matrix, and the third displays the mean absolute deviation for the intrinsic frequencies.  In panels (a-c) reconstruction of the coupling function and adjacency matrix fail regardless of the number of perturbations due to the small perturbation size.  In the remaining panels, reconstruction is successful once there are a sufficient number of perturbations. Here the oscillators are nearly identical: $\sigma=0.0001$ and with initial condition $\theta_k=0$ for all $k$. }
\label{fig:Repeat}
 \end{figure}

As such we investigated a method for using small perturbations to introduce brief transients into the dynamics to allow for successful model reconstruction. To test this, we used nearly identical oscillators with frequency standard deviation $\sigma=0.0001$, which causes the system to quickly converge to a synchronized state for almost all initial conditions.  We then initialized $\bm{\theta}(0)=\bm{0}$ so that the system begins from perfect synchrony.  Then, at times $kt_{max}/N_{pert}$ for $k=1,2,\ldots N_{pert}$ we added a phase perturbation to a subset of the oscillators. Each perturbation causes some of the oscillators to briefly become desynchronized. In this way, the total number of observed phases remains constant, but the fraction of those observations that occur during transient dynamics is proportional to the number of perturbations, $N_{pert}$. The observed phases during these transients provide meaningful data about the structure of the network.  \Fref{fig:Repeat} demonstrates that, as expected, with too few or too small perturbations, model reconstruction is unsuccessful.  However, as the number of perturbations $N_{pert}$ increases, the accuracy of the estimated coupling function, adjacency matrix, and the oscillator frequencies improves. 

 % combined figure: # of repeats

We explored two methods for selecting which oscillators to perturb: fixed subsets, in which the subset of perturbed oscillators was selected at the beginning of the experiment and these same oscillators were perturbed repeatedly, and random subsets, in which a random subset of the system was selected for each perturbation. Although a perturbation to the phase of a particular oscillator does propagate to the phases of neighboring oscillators through the coupling terms, those perturbations decay quickly and are therefore most effective for revealing the local structure of the network.  As such, although perturbations to a fixed subset might be more practical in a physical experiment, one must typically perturb a larger fraction of the system to obtain comparable performance to that which is obtained with perturbations to random subsets.

We also considered two types of phase perturbations: phase resets, in which selected oscillators had their phases reset to 0, and phase shifts, in which selected oscillators had their phases modified by adding a random shift  $\eta_{pert}\sim \mathcal{N}(0,\sigma^2_{pert})$. Phase resets may be more feasible from an experimental perspective, but have the drawback of preserving the mutual synchrony of the subset of oscillators that are perturbed.  As such one will typically need to use random subsets in tandem with phase resets in order to be able to resolve the connections between the perturbed oscillators.

In \fref{fig:Repeat}(a-c), we used phase shifts with $\sigma_{pert}=0.01$ to a fixed subset of size 1, i.e. a single oscillator out of $10$. This gives poor reconstruction regardless of the number of perturbations due to the small size of the perturbation. On the other hand, in  \fref{fig:Repeat}(d-f) we perturb a fixed subset (3 out of 10 oscillators) with random phase shifts with $\sigma_{pert}=10$ \footnote{This is virtually indistinguishable from uniform perturbations $X\sim \mathcal{U}[-\pi,\pi]$}.  In this case, the perturbations affect a sufficiently large proportion of the oscillators and performance begins to improve once there are 5 or more perturbations. The results in \fref{fig:Repeat}(g-i) illustrate that one only needs to perturb 1 out of 10 oscillators to obtain similar performance when the oscillators selected are chosen randomly. In \fref{fig:Repeat}(j-l), we show the results using phase resets to random subsets of 3 oscillators. Again, performance begins to improve dramatically once 5 or more perturbations are used. The tables summarizing the results of these perturbation strategies are provided in \ref{app:tables}.

\subsection{Comparison with reference \cite{Pik2018}} \label{sec:pikresults}
 
The amount of transient data necessary for reconstruction is also useful in comparing our method to previous approaches. As discussed in \sref{sec:inv}, Pikovsky proposes an alternative method where distinct coupling functions are considered for the interaction of each pair of oscillators \cite{Pik2018}. This ensures that the system of equations in the optimization is linear, however it also increases the number of unknown coefficients significantly. Our approach relies on the assumption that the same coupling function $\Gamma$ is used for all pairs of oscillators.  This is a reasonable approximation for physical systems when the physical mechanisms governing the interactions between oscillators are the same. Pikovsky \cite{Pik2018} uses a more general model in which the coupling functions may be different for each pair of oscillators.

\begin{figure}[t!]
  \centering
 \sidesubfloat[]{\includegraphics[width=0.45\textwidth]{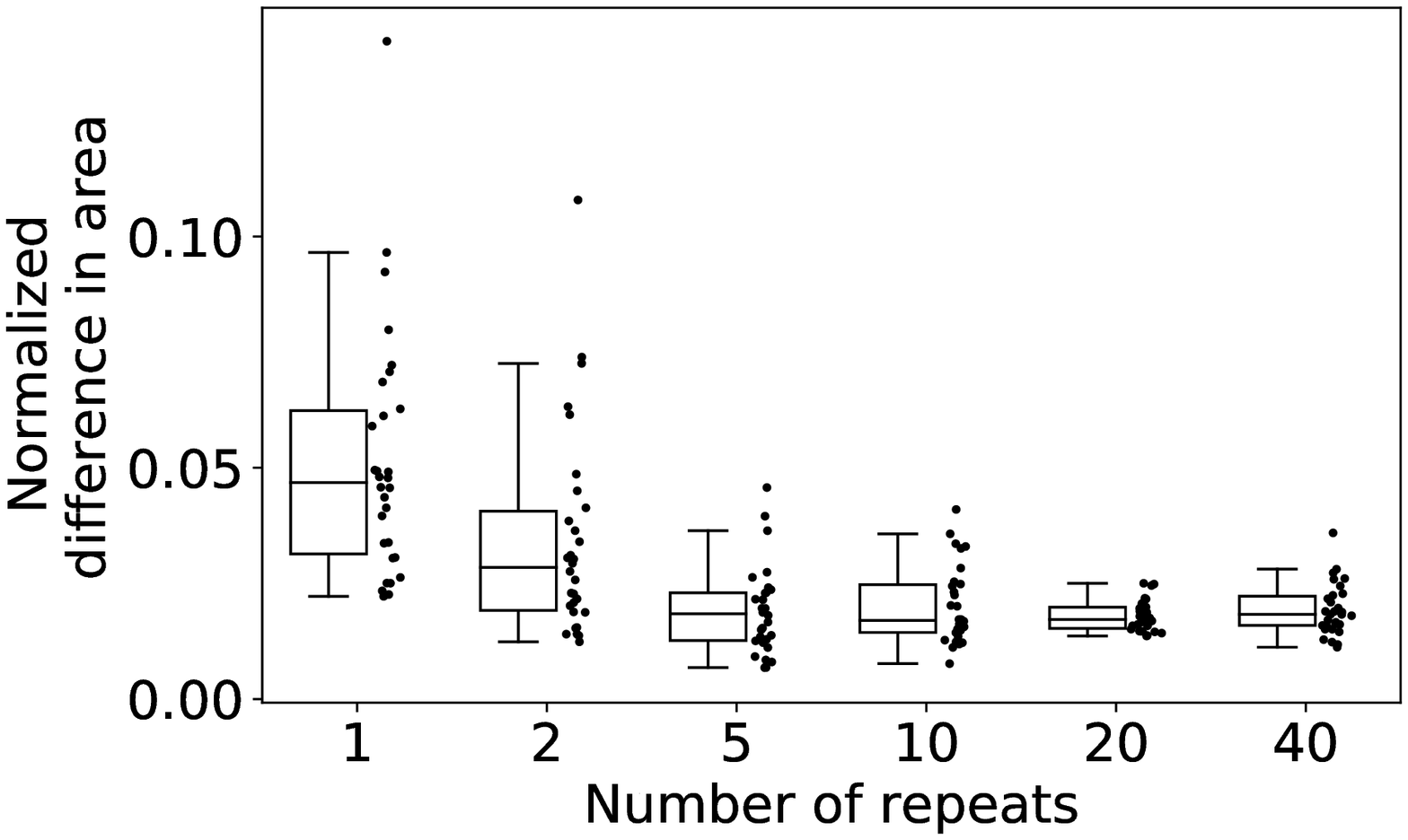}}
  \sidesubfloat[]{
  \includegraphics[width=0.45\textwidth]{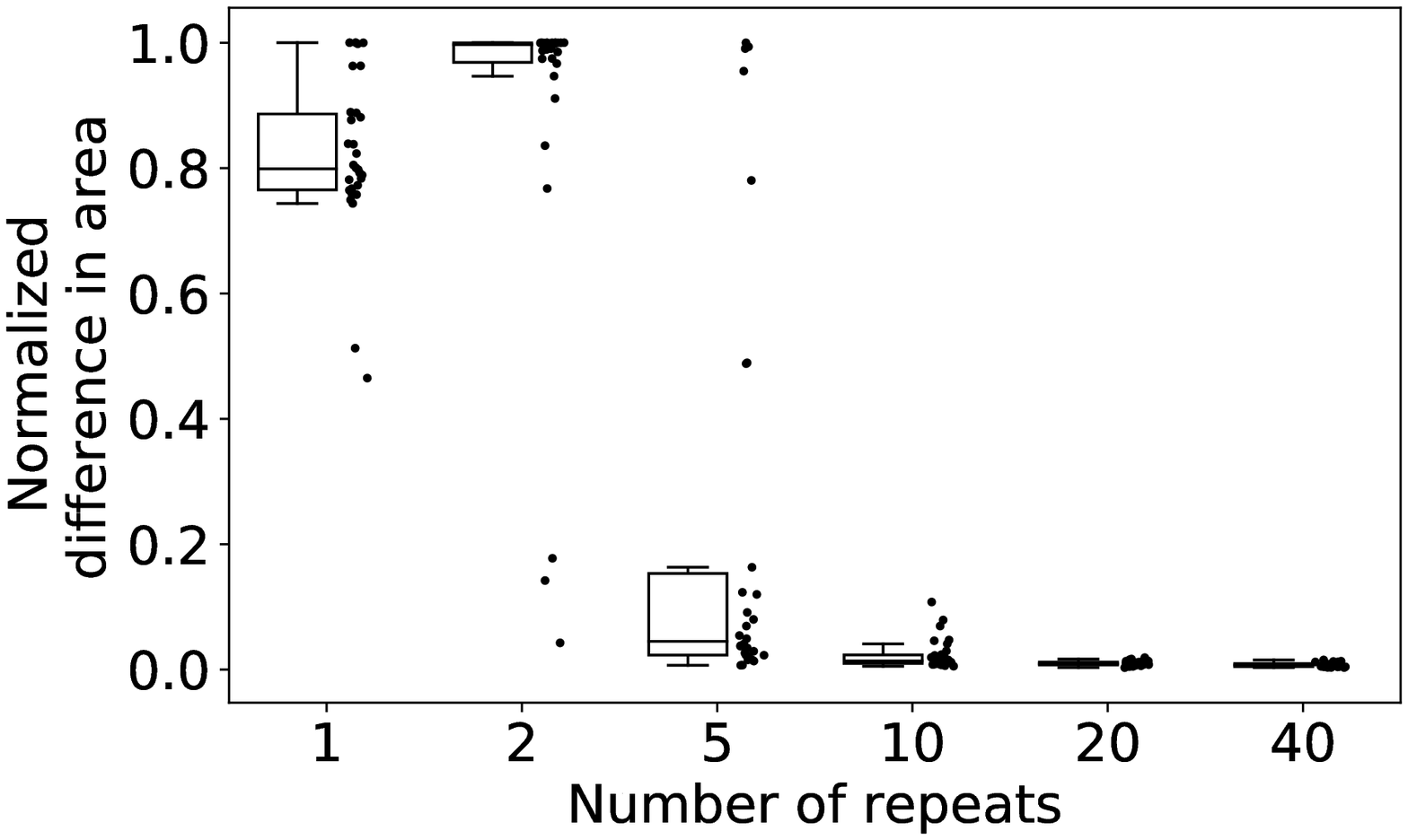}}
  \hfill
 \sidesubfloat[]{\includegraphics[width=0.45\textwidth]{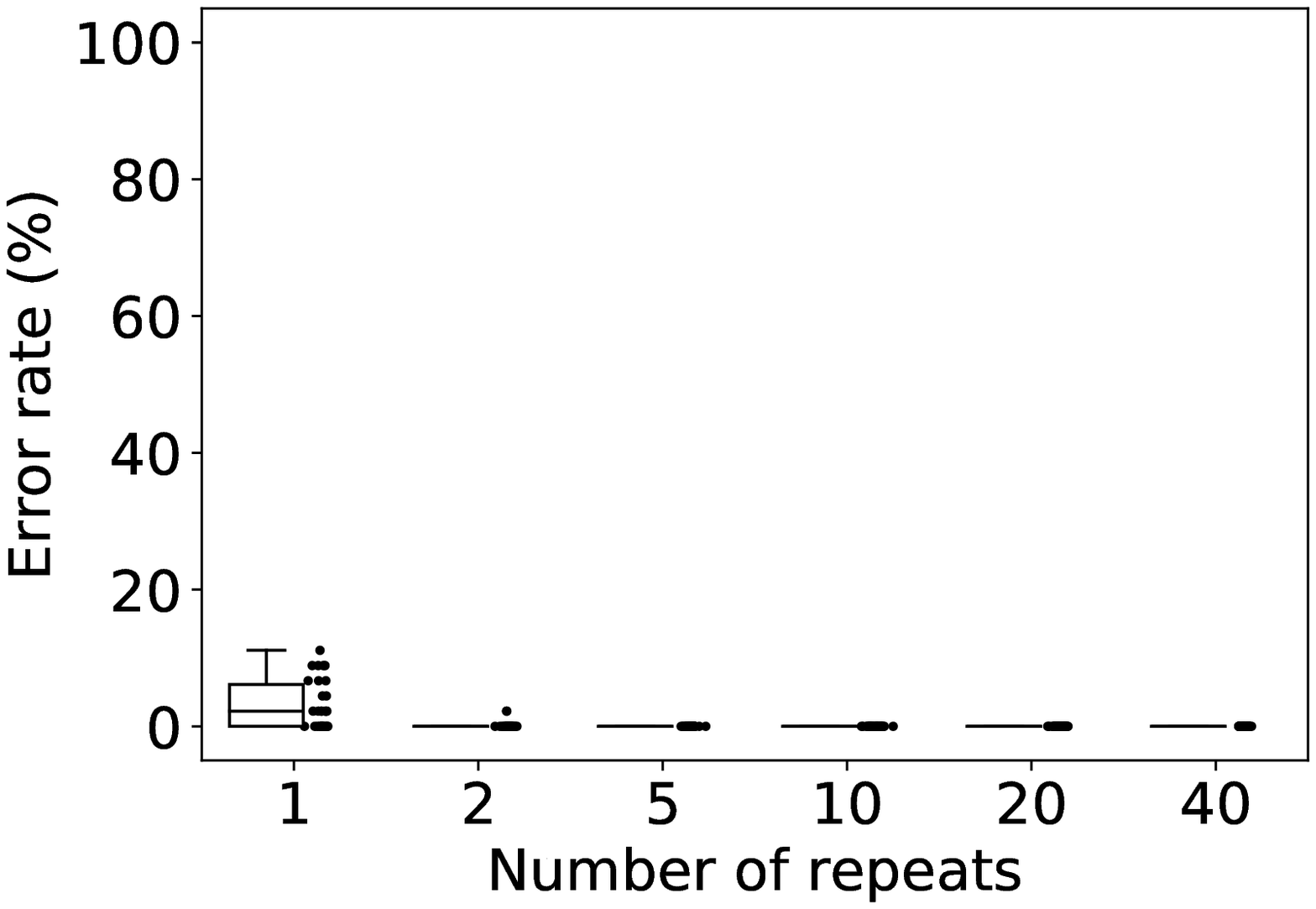}}
 \sidesubfloat[]{
 \includegraphics[width=0.45\textwidth]{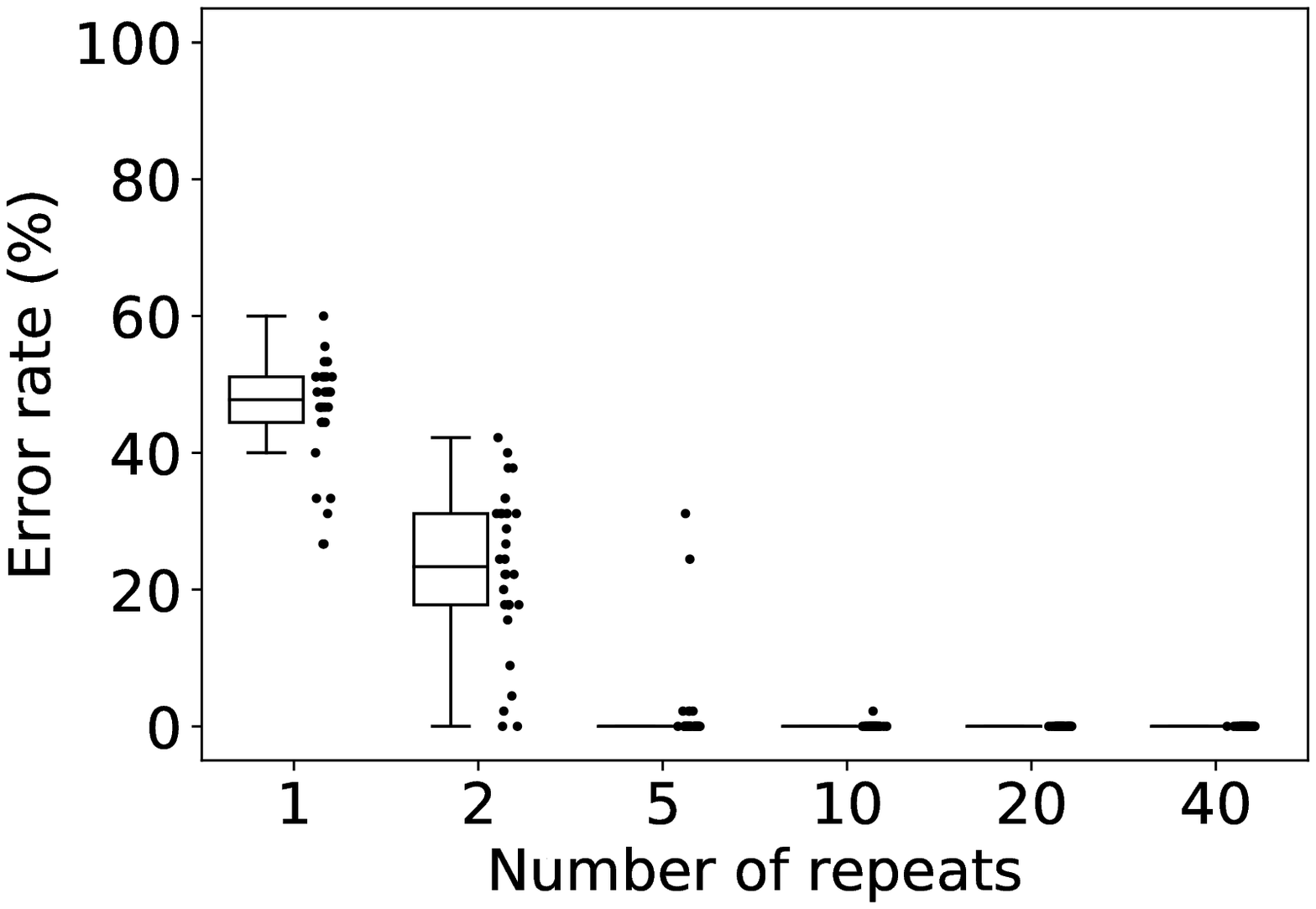}}
 \caption{Comparison of our method (left column) with the method described in reference \cite{Pik2018} (right column), for different numbers of restarts. Panels (a) and (b) display the normalized difference in area of the coupling function. For the coupling function reconstruction, our method performs well with a small number of restarts while Pikovsky's method requires a relatively larger number of restarts. Panels (c) and (d) display the error rate of the adjacency matrix. Again, both methods perform well with 10 or more restarts. However, our method provides more accurate reconstructions when the number of restarts is five or less.}
\label{fig:sweep4}
 \end{figure}

\Fref{fig:sweep4} shows that when the coupling functions are identical, our method provides more accurate reconstructions of both the network topology and the coupling function with smaller amounts of transient data. As the amount of transient data increases, both methods ultimately achieve perfect network reconstruction while the approach in \cite{Pik2018} ultimately obtains slightly better estimates for the coupling function. Therefore, our approach would be preferred under circumstances where the amount of data is limited and when the coupling functions are nearly identical.

\section{Discussion and conclusions} \label{sec:conclusions}
In this paper, we have designed a method for reconstructing models of coupled oscillator networks including both the network connections and the intrinsic oscillator properties. We hope analysts might investigate (non)convexity of our penalty function and properties of  minimizers. These issues aside, after testing our method with many different parameters, we conclude that our algorithm can successfully infer the model.

A common challenge that our method and many related ones encounter is that the procedure fails if the system synchronizes too quickly. One remedy is to have a sufficient number of observations during the desynchronized transients. As we demonstrate, one can ensure that this data is available by repeating experiments with multiple initial conditions or by adjusting the model parameters to inhibit synchronization by instituting large frequency variability ($\sigma$) or dynamic noise.  
An alternative remedy is to move the system away from synchrony using perturbations, preferably ones that are physically realizable. The introduction of these perturbations provides useful transient data. By perturbing a sufficiently large subset of oscillators with a large enough change, we are able to infer the model accurately. Our hope is that this method will be adopted by experimentalists and used with experimental data to aid with the construction of interpretable models for the dynamics of networks of coupled oscillators. 

Although the numerical experiments outlined here involve Erd\"os-R\'enyi networks, the method can be applied more generally to a broad class of networks.  Preliminary testing suggests that similar results can be obtained for other topologies such as star, small-world, scale-free networks, and clique networks.  

It is also straightforward to extend this approach to other models for coupled phase oscillators such as the Winfree model \cite{winfree2001geometry} or even more general oscillator models such as the Stuart-Landau model \cite{nakagawa1993collective} in which both phase and amplitude variations are permitted. Indeed, any system where the unknown functions are periodic can be represented using our technique.

For models containing unknown functions that are not periodic such as the Hodgkin-Huxley model \cite{hodgkin1952quantitative}, a Fourier series representation is not possible. In these cases, one could represent the coupling functions using feed-forward neural networks, which are capable of representing continuous functions to arbitrary accuracy using a finite number of parameters \cite{Hor1991}. Given a sufficiently rich data set, one could still use our approach with back-propagation to learn the structure of the neural network approximation for the coupling function.

\ack This material is based upon work supported by the  Mathematics  Research  Communities  of  the  American Mathematical Society, under National Science Foundation grant DMS-1321794. The project was initiated during the Mathematics Research Community (MRC) on Agent-based Modeling in Biological and Social Systems (2018). A follow-up visit was also supported by a collaboration travel grant from the AMS MRC program. MVC was supported by The Ohio State University President's Postdoctoral Scholars Program and by the Mathematical Biosciences Institute at The Ohio State University. MJP is supported by Hillsdale College. CMT is supported by National Science Foundation grant DMS-1813752 and by Williams College. BX is supported by the Robert and Sara Lumpkins Endowment for Postdoctoral Fellows in Applied and Computational Math and Statistics at the University of Notre Dame. We are grateful to Henry Adams, Kelsey Houston-Edwards, and Lori Ziegelmeier for contributions during the formative stages of this work.

\newpage 
\appendix
\setcounter{section}{0}

\section{Coupling functions}\label{app:coup_funcs}
The coupling functions investigated are described in \tref{tab:coup}. The first three classes of functions were selected due to their use in popular coupled oscillator models.  The last was an example of a generic coupling function with higher order harmonics that was tested to verify that reconstruction of the adjacency matrix is possible even with an imperfect approximation to the coupling function.

\begin{table}[h!]
\caption{Coupling functions}
\label{tab:coup}
\begin{tabular}{@{}l*{15}{l}}
\br
Name & Function & Reference \\
\mr
Kuramoto & $\sin x$ & \cite{Kur1975}\\
Kuramoto-Sakaguchi & $\sin(x-0.1)$ & \cite{SakKur1986}\\[10pt]
Hodgkin-Huxley & 
$\begin{array}{r@{}l@{}} 
0.383&+1.379\sin(x+3.93)\\
&+0.568\sin(2x+0.11)+0.154\sin(3x+2.387)
\end{array}$
 &  \cite{HanMatMeu1993}\\[10pt]
Square Wave & $\displaystyle\frac{\sin(x-\pi/4)}{|\sin(x-\pi/4)|}$ & N/A\\
\br
\end{tabular}
\end{table}

\section{Evaluation metrics for classification}\label{app:perf_metrics}
The \textit{$F_1$ score} is defined as \[\left( \frac{ \mathrm{precision}^{-1} + \mathrm{recall}^{-1}}{2} \right)^{-1}  = 2 \times \frac{ \mathrm{precision} \times \mathrm{recall}}{\mathrm{precision} + \mathrm{recall}}.\]
Here, $\mathrm{precision}$ is the fraction of true positives among all inferred positives, while $\mathrm{recall}$ is the fraction of true positives among all positives. The $F_1$ score is a value between $0$ and $1$; when reporting the error rate for the reconstructed adjacency matrix, we use the threshold $\epsilon=\epsilon_{max}$ that corresponds to the largest $F_1$ score. We also determine the interval of thresholds that yield $F_1$ scores within 90\% of this largest value. The width of this interval is reported in the tables in \ref{app:tables} as ``Interval width $(>90\%)$".

The \textit{error rate} is defined as the percentage of entries in $A$ that are incorrectly classified. We report the error rate for the optimal threshold $\epsilon$ discussed in \sref{sec:postproc}.

The \textit{ROC} curve, or receiver operating characteristic, is a parametric curve that uses the classification threshold as a parameter and uses the true positive rate and the false positive rate  as the the variables \cite{Faw2006}. The \textit{area under the ROC curve} measures the quality of a classifier independent any particular threshold.  A value of $1/2$ is consistent with blind guesses, and a value of $1$ indicates a perfect classifier since it implies the existence of a threshold for which the rate of true positives is $1$ and the rate of false positives is $0$. Our ROC curves were generated using the scikit-learn package in Python \cite{PedVarGra2011}.

\section{Tables of results}\label{app:tables}

Here we include several tables which report the averaged numerical results of the experimental parameter sweeps described in \sref{sec:experiments} and reported on in \sref{sec:results}. The performance metrics used are defined in \ref{app:perf_metrics}. Values given in the tables represent the mean plus or minus one standard deviation of the corresponding performance metric over the 30 trials run at that parameter value. We note that some of the distributions, such as the error rate (which is nonnegative by definition), are skewed.

\begin{table}[h!]
\footnotesize
\begin{tabular}{llllllll}
\br
Param   & Value  & \begin{tabular}{@{}l@{}} Normalized \\ difference in area \end{tabular} & \begin{tabular}{@{}l@{}}Mean absolute\\ deviation\end{tabular} & Error rate           & \begin{tabular}{@{}l@{}} Area under\\ ROC curve\end{tabular} & \begin{tabular}{@{}l@{}} Interval \\width $(>90\%)$  \end{tabular}\\
\mr
$\Gamma$ & Kuramoto                 & 0.0175 $\pm$ 0.0075 & 0.004 $\pm$ 0.001      & 0.0 $\pm$ 0.0 & 1.0$\pm$ 0.0         & 0.4338                 \\
$\Gamma$ & Kuramoto-Sakaguchi & 0.0169 $\pm$ 0.0054 & 0.005 $\pm$ 0.002      & 0.0 $\pm$ 0.0 & 1.0$\pm$ 0.0         & 0.3924                 \\
$\Gamma$ &  Hodgkin-Huxley           & 0.0343 $\pm$ 0.0159 & 0.011 $\pm$ 0.004     & 0.0 $\pm$ 0.0 & 1.0$\pm$ 0.0         & 0.7054                 \\
$\Gamma$ &  Square wave              & 0.1518 $\pm$ 0.0044 & 0.006 $\pm$ 0.002      & 0.0 $\pm$ 0.0 & 1.0$\pm$ 0.0         & 0.4059\\
\br
\end{tabular}
\caption{Sweep through the coupling function $\Gamma$}
\label{tab:coupling_fcn}
\end{table}

\begin{table}[h!]
\footnotesize
\begin{tabular}{lllllll}
\br
Param   & Value  & \begin{tabular}{@{}l@{}} Normalized \\ difference in area \end{tabular} & \begin{tabular}{@{}l@{}}Mean absolute\\ deviation\end{tabular} & Error rate           & \begin{tabular}{@{}l@{}} Area under\\ ROC curve\end{tabular} & \begin{tabular}{@{}l@{}} Interval \\width $(>90\%)$  \end{tabular}\\
\mr
$N$       & 5     & 0.026 $\pm$ 0.0125  & 0.0052 $\pm$ 0.003      & 0.0 $\pm$ 0.0       & 1.0$\pm$ 0.0         & 0.1193                     \\
$N$   & 10    & 0.0164 $\pm$ 0.007  & 0.0047 $\pm$ 0.0031     & 0.0 $\pm$ 0.0       & 1.0$\pm$ 0.0         & 0.4057                     \\
$N$   & 20    & 0.015 $\pm$ 0.0057  & 0.0046 $\pm$ 0.0016     & 0.140 $\pm$ 0.601 & 0.9997$\pm$ 0.0016   & 0.3041                     \\
$N$   & 40    & 0.0189 $\pm$ 0.0104 & 0.0045 $\pm$ 0.0011     & 1.231 $\pm$ 1.018 & 0.9946$\pm$ 0.0053   & 0.223  \\              \br 
\end{tabular}
\caption{Sweep of the number of oscillators $N$}
\label{tab:num_osc}
\end{table}

\begin{table}[h!]
\footnotesize
\begin{tabular}{lllllll}
\br
Param   & Value  & \begin{tabular}{@{}l@{}} Normalized \\ difference in area \end{tabular} & \begin{tabular}{@{}l@{}}Mean absolute\\ deviation\end{tabular} & Error rate           & \begin{tabular}{@{}l@{}} Area under\\ ROC curve\end{tabular} & \begin{tabular}{@{}l@{}} Interval \\width $(>90\%)$  \end{tabular}\\
\mr
$\sigma$ & 0.01  & 0.0198 $\pm$ 0.0052 & 0.0022 $\pm$ 0.0008     & 0.0 $\pm$ 0.0 & 1.0$\pm$ 0.0         & 0.8142                   \\
$\sigma$ & 0.1   & 0.0175 $\pm$ 0.0057 & 0.004 $\pm$ 0.0014      & 0.0 $\pm$ 0.0 & 1.0$\pm$ 0.0         & 0.8095                   \\
$\sigma$ & 1     & 0.0161 $\pm$ 0.0049 & 0.0042 $\pm$ 0.0021     & 0.0 $\pm$ 0.0 & 1.0$\pm$ 0.0         & 0.7975     \\
\br
\end{tabular}
\caption{Sweep of the standard deviation of the oscillator frequencies $\sigma$}
\label{tab:sigma_freq}
\end{table}

\begin{table}[h!]
\footnotesize
\begin{tabular}{lllllll}
\br
Param   & Value  & \begin{tabular}{@{}l@{}} Normalized \\ difference in area \end{tabular} & \begin{tabular}{@{}l@{}}Mean absolute\\ deviation\end{tabular} & Error rate           & \begin{tabular}{@{}l@{}} Area under\\ ROC curve\end{tabular} & \begin{tabular}{@{}l@{}} Interval \\width $(>90\%)$  \end{tabular}\\
\mr
noise & 0      & 0.0193 $\pm$ 0.0098 & 0.0048 $\pm$ 0.002      & 0.0 $\pm$ 0.0        & 1.0$\pm$ 0.0         & 0.4204                     \\
noise & 1e-05  & 0.0202 $\pm$ 0.0108 & 0.0053 $\pm$ 0.0024     & 0.0 $\pm$ 0.0        & 1.0$\pm$ 0.0         & 0.3845                     \\
noise & 0.0001 & 0.018 $\pm$ 0.0076  & 0.0048 $\pm$ 0.0023     & 0.0 $\pm$ 0.0        & 1.0$\pm$ 0.0         & 0.4365                     \\
noise & 0.001  & 0.0186 $\pm$ 0.0081 & 0.0053 $\pm$ 0.002      & 0.0 $\pm$ 0.0        & 1.0$\pm$ 0.0         & 0.3956                     \\
noise & 0.01   & 0.0186 $\pm$ 0.0079 & 0.0052 $\pm$ 0.0025     & 0.0 $\pm$ 0.0        & 1.0$\pm$ 0.0         & 0.4135                     \\
noise & 0.1    & 0.0385 $\pm$ 0.0138 & 0.0086 $\pm$ 0.0021     & 0.0 $\pm$ 0.0        & 1.0$\pm$ 0.0         & 0.3544                     \\
noise & 1      & 0.831 $\pm$ 0.1343  & 0.118 $\pm$ 0.0629      & 46.296 $\pm$ 8.579 & 0.5122$\pm$ 0.0741   & 0.0472   \\          \br       
\end{tabular}
\caption{Sweep of the level of observation noise}
\label{tab:noise_level}
\end{table}

\begin{table}[h!]
\footnotesize
\begin{tabular}{lllllll}
\br
Param   & Value  & \begin{tabular}{@{}l@{}} Normalized \\ difference in area \end{tabular} & \begin{tabular}{@{}l@{}}Mean absolute\\ deviation\end{tabular} & Error rate           & \begin{tabular}{@{}l@{}} Area under\\ ROC curve\end{tabular} & \begin{tabular}{@{}l@{}} Interval \\width $(>90\%)$  \end{tabular}\\
\mr
dyn\_noise & 0      & 0.0211 $\pm$ 0.0067 & 0.0055 $\pm$ 0.0018     & 0.0 $\pm$ 0.0        & 1.0$\pm$ 0.0         & 0.3814                     \\
dyn\_noise & 1e-05  & 0.0117 $\pm$ 0.0069 & 0.0032 $\pm$ 0.0023     & 0.0 $\pm$ 0.0        & 1.0$\pm$ 0.0         & 0.4048                     \\
dyn\_noise & 0.0001 & 0.0146 $\pm$ 0.0049 & 0.0032 $\pm$ 0.0013     & 0.0 $\pm$ 0.0        & 1.0$\pm$ 0.0         & 0.411                      \\
dyn\_noise & 0.001  & 0.0241 $\pm$ 0.0056 & 0.0049 $\pm$ 0.0017     & 0.0 $\pm$ 0.0        & 1.0$\pm$ 0.0         & 0.4517                     \\
dyn\_noise & 0.01   & 0.0633 $\pm$ 0.0189 & 0.0128 $\pm$ 0.0034     & 0.370 $\pm$ 1.025  & 0.9982$\pm$ 0.0072   & 0.3282                     \\
dyn\_noise & 0.1    & 0.2136 $\pm$ 0.065  & 0.0312 $\pm$ 0.0055     & 14.148 $\pm$ 5.003 & 0.8903$\pm$ 0.0496   & 0.2224                     \\
dyn\_noise & 1      & 0.9716 $\pm$ 0.0399 & 0.1552 $\pm$ 0.0971     & 42.444 $\pm$ 8.953 & 0.5177$\pm$ 0.1004   & 0.0406 \\       
\br
\end{tabular}
\caption{Sweep of the level of noise in system dynamics}
\label{tab:dynamic_noise_level}
\end{table}

\begin{table}[h!]
\footnotesize
\begin{tabular}{llllllll}
\br
Param   & Value  & \begin{tabular}{@{}l@{}} Normalized \\ difference in area \end{tabular} & \begin{tabular}{@{}l@{}}Mean absolute\\ deviation\end{tabular} & Error rate           & \begin{tabular}{@{}l@{}} Area under\\ ROC curve\end{tabular} & \begin{tabular}{@{}l@{}} Interval \\width $(>90\%)$  \end{tabular}\\
\mr
$p$  & 0.1   & 0.1199 $\pm$ 0.2314 & 0.0054 $\pm$ 0.0047     & 3.111 $\pm$ 14.97 & 0.9688$\pm$ 0.1548   & 0.3387                 \\
$p$  & 0.2   & 0.023 $\pm$ 0.0098  & 0.004 $\pm$ 0.0022      & 0.074 $\pm$ 0.41  & 0.9999$\pm$ 0.0006   & 0.5222                 \\
$p$  & 0.3   & 0.0186 $\pm$ 0.0073 & 0.0047 $\pm$ 0.0025     & 0.0 $\pm$ 0.0        & 1.0$\pm$ 0.0         & 0.5175                 \\
$p$  & 0.4   & 0.0207 $\pm$ 0.0089 & 0.0047 $\pm$ 0.0022     & 0.0 $\pm$ 0.0        & 1.0$\pm$ 0.0         & 0.5167                 \\
$p$  & 0.5   & 0.0189 $\pm$ 0.0097 & 0.0051 $\pm$ 0.0024     & 0.0 $\pm$ 0.0        & 1.0$\pm$ 0.0         & 0.4669                 \\
$p$  & 0.6   & 0.0183 $\pm$ 0.0072 & 0.006 $\pm$ 0.0019      & 0.0 $\pm$ 0.0        & 1.0$\pm$ 0.0         & 0.349                  \\
$p$  & 0.7   & 0.0169 $\pm$ 0.0071 & 0.005 $\pm$ 0.0025      & 0.0 $\pm$ 0.0        & 1.0$\pm$ 0.0         & 0.3227                 \\
$p$  & 0.8   & 0.0194 $\pm$ 0.0077 & 0.0057 $\pm$ 0.0026     & 0.0 $\pm$ 0.0        & 1.0$\pm$ 0.0         & 0.2725                 \\
$p$  & 0.9   & 0.0289 $\pm$ 0.0271 & 0.0077 $\pm$ 0.0056     & 0.0 $\pm$ 0.0        & 1.0$\pm$ 0.0         & 0.2739 \\  
\br
\end{tabular}
\caption{Sweep of the network connectivity parameter $p$}
\label{tab:p_erdos}
\end{table}

\begin{table}[h!]
\footnotesize
\begin{tabular}{lllllll}
\br
Param   & Value  & \begin{tabular}{@{}l@{}} Normalized \\ difference in area \end{tabular} & \begin{tabular}{@{}l@{}}Mean absolute\\ deviation\end{tabular} & Error rate           & \begin{tabular}{@{}l@{}} Area under\\ ROC curve\end{tabular} & \begin{tabular}{@{}l@{}} Interval \\width $(>90\%)$  \end{tabular}\\
\mr
$t_{\max}$     & 2     & 0.0204 $\pm$ 0.011  & 0.0042 $\pm$ 0.002      & 0.0 $\pm$ 0.0 & 1.0$\pm$ 0.0         & 0.8225                     \\
$t_{\max}$       & 5     & 0.0185 $\pm$ 0.0064 & 0.0047 $\pm$ 0.0022     & 0.0 $\pm$ 0.0 & 1.0$\pm$ 0.0         & 0.8186                     \\
$t_{\max}$      & 10    & 0.0193 $\pm$ 0.0066 & 0.004 $\pm$ 0.0013      & 0.0 $\pm$ 0.0 & 1.0$\pm$ 0.0         & 0.8175                     \\
$t_{\max}$      & 20    & 0.0192 $\pm$ 0.0053 & 0.0041 $\pm$ 0.0013     & 0.0 $\pm$ 0.0 & 1.0$\pm$ 0.0         & 0.565                      \\
$t_{\max}$      & 50    & 0.0191 $\pm$ 0.0093 & 0.0064 $\pm$ 0.0038     & 0.0 $\pm$ 0.0 & 1.0$\pm$ 0.0         & 0.2964                     \\
\br
\end{tabular}
\caption{Sweep of the simulation time $t_{\max}$ (duration of each transient)}
\label{tab:tmax}
\end{table}

\begin{table}[h!]
\footnotesize
\begin{tabular}{lllllll}
\br
Param   & Value  & \begin{tabular}{@{}l@{}} Normalized \\ difference in area \end{tabular} & \begin{tabular}{@{}l@{}}Mean absolute\\ deviation\end{tabular} & Error rate           & \begin{tabular}{@{}l@{}} Area under\\ ROC curve\end{tabular} & \begin{tabular}{@{}l@{}} Interval \\width $(>90\%)$  \end{tabular}\\
\mr
$N_{res}$ & 1     & 0.0513 $\pm$ 0.0266 & 0.0237 $\pm$ 0.0127     & 3.111 $\pm$ 3.531 & 0.9815$\pm$ 0.027    & 0.447                      \\
$N_{res}$ & 2     & 0.0343 $\pm$ 0.0222 & 0.0082 $\pm$ 0.007      & 0.148 $\pm$ 0.564 & 0.9986$\pm$ 0.0067   & 0.6943                     \\
$N_{res}$ & 5     & 0.019 $\pm$ 0.0094  & 0.0039 $\pm$ 0.002      & 0.0 $\pm$ 0.0       & 1.0$\pm$ 0.0         & 0.806                      \\
$N_{res}$ & 10    & 0.0201 $\pm$ 0.0084 & 0.0042 $\pm$ 0.0028     & 0.0 $\pm$ 0.0       & 1.0$\pm$ 0.0         & 0.808                      \\
$N_{res}$ & 20    & 0.018 $\pm$ 0.0033  & 0.0042 $\pm$ 0.0017     & 0.0 $\pm$ 0.0       & 1.0$\pm$ 0.0         & 0.8295                     \\
$N_{res}$ & 40    & 0.0192 $\pm$ 0.0055 & 0.0041 $\pm$ 0.0016     & 0.0 $\pm$ 0.0       & 1.0$\pm$ 0.0         & 0.8178    \\
\br
\end{tabular}
 \caption{Sweep through the number of transients observed for a simulation that re-initializes all oscillator phases from $\mathcal{U}[0,2\pi]$.}
\label{tab:full_refresh}
 \end{table}

\begin{table}[h!]
\footnotesize
\begin{tabular}{lllllll}
\br
Param   & Value  & \begin{tabular}{@{}l@{}} Normalized \\ difference in area \end{tabular} & \begin{tabular}{@{}l@{}}Mean absolute\\ deviation\end{tabular} & Error rate           & \begin{tabular}{@{}l@{}} Area under\\ ROC curve\end{tabular} & \begin{tabular}{@{}l@{}} Interval \\width $(>90\%)$  \end{tabular}\\
\mr
$N_{pert}$ & 1     & 0.9318 $\pm$ 0.082  & 0.118 $\pm$ 0.0719      & 48.667 $\pm$ 7.459 & 0.2934$\pm$ 0.0999   & 0.0252                     \\
$N_{pert}$ & 2     & 0.9054 $\pm$ 0.1176 & 0.3284 $\pm$ 0.0747     & 49.778 $\pm$ 6.920 & 0.2904$\pm$ 0.0783   & 0.0306                     \\
$N_{pert}$ & 5     & 0.7928 $\pm$ 0.1568 & 0.3853 $\pm$ 0.1243     & 51.852 $\pm$ 8.702 & 0.3281$\pm$ 0.0895   & 0.0688                     \\
$N_{pert}$ & 10    & 0.7828 $\pm$ 0.1698 & 0.2566 $\pm$ 0.1579     & 50.593 $\pm$ 5.110 & 0.3843$\pm$ 0.0697   & 0.1403                     \\
$N_{pert}$ & 20    & 0.9818 $\pm$ 0.0083 & 0.0024 $\pm$ 0.0011     & 43.630 $\pm$ 9.383 & 0.5153$\pm$ 0.0959   & 0.1026                     \\
$N_{pert}$ & 40    & 0.931 $\pm$ 0.0234  & 0.0019 $\pm$ 0.0007     & 47.556 $\pm$ 9.403 & 0.4909$\pm$ 0.0935   & 0.0931  \\        
\br
\end{tabular}
\caption{Sweep through the number of transients observed for a simulation that selects a fixed oscillator and adds a random Gaussian perturbation to its phase with standard deviation $0.01$.}
\label{tab:1fixed_p01}
\end{table}

\begin{table}[h!]
\footnotesize
\begin{tabular}{lllllll}
\br
Param   & Value  & \begin{tabular}{@{}l@{}} Normalized \\ difference in area \end{tabular} & \begin{tabular}{@{}l@{}}Mean absolute\\ deviation\end{tabular} & Error rate           & \begin{tabular}{@{}l@{}} Area under\\ ROC curve\end{tabular} & \begin{tabular}{@{}l@{}} Interval \\width $(>90\%)$  \end{tabular}\\
\mr
$N_{pert}$ & 1     & 0.9193 $\pm$ 0.1111 & 0.1161 $\pm$ 0.0643     & 49.778 $\pm$ 7.042  & 0.3189$\pm$ 0.1051   & 0.0298                     \\
$N_{pert}$ & 2     & 0.7919 $\pm$ 0.2112 & 0.343 $\pm$ 0.0724      & 49.852 $\pm$ 6.331  & 0.279$\pm$ 0.0949    & 0.0409                     \\
$N_{pert}$ & 5     & 0.4298 $\pm$ 0.1535 & 0.3328 $\pm$ 0.0691     & 39.778 $\pm$ 11.032 & 0.5271$\pm$ 0.1261   & 0.2041                     \\
$N_{pert}$ & 10    & 0.3211 $\pm$ 0.1187 & 0.1697 $\pm$ 0.0769     & 22.741 $\pm$ 10.975 & 0.7388$\pm$ 0.0974   & 0.2111                     \\
$N_{pert}$ & 20    & 0.1311 $\pm$ 0.0931 & 0.0288 $\pm$ 0.0302     & 4.667 $\pm$ 5.296   & 0.9536$\pm$ 0.0601   & 0.3584                     \\
$N_{pert}$ & 40    & 0.0447 $\pm$ 0.0311 & 0.0036 $\pm$ 0.0026     & 0.222 $\pm$ 0.679   & 0.9988$\pm$ 0.0057   & 0.4824 \\
\br
\end{tabular}
\caption{Sweep through the number of transients observed for a simulation that selects a random oscillator and adds a random Gaussian perturbation to its phase with standard deviation $10$.}
\label{tab:1rand_10}
\end{table}

\begin{table}[h!]
\footnotesize
\begin{tabular}{lllllll}
\br
Param   & Value  & \begin{tabular}{@{}l@{}} Normalized \\ difference in area \end{tabular} & \begin{tabular}{@{}l@{}}Mean absolute\\ deviation\end{tabular} & Error rate           & \begin{tabular}{@{}l@{}} Area under\\ ROC curve\end{tabular} & \begin{tabular}{@{}l@{}} Interval \\width $(>90\%)$  \end{tabular}\\
\mr
$N_{pert}$ & 1     & 0.8851 $\pm$ 0.1332 & 0.1203 $\pm$ 0.0716     & 50.0 $\pm$ 5.277    & 0.316$\pm$ 0.1035    & 0.0293                     \\
$N_{pert}$ & 2     & 0.7678 $\pm$ 0.215  & 0.2857 $\pm$ 0.0916     & 50.963 $\pm$ 6.879  & 0.315$\pm$ 0.0915    & 0.0326                     \\
$N_{pert}$ & 5     & 0.286 $\pm$ 0.0821  & 0.1973 $\pm$ 0.0625     & 30.963 $\pm$ 9.096  & 0.6136$\pm$ 0.1118   & 0.1767                     \\
$N_{pert}$ & 10    & 0.1567 $\pm$ 0.0624 & 0.0788 $\pm$ 0.0609     & 15.704 $\pm$ 4.246 & 0.7659$\pm$ 0.0863   & 0.3186                     \\
$N_{pert}$ & 20    & 0.0783 $\pm$ 0.056  & 0.0268 $\pm$ 0.06       & 10.815 $\pm$ 4.943 & 0.8699$\pm$ 0.0777   & 0.4238                     \\
$N_{pert}$ & 40    & 0.0262 $\pm$ 0.0191 & 0.0041 $\pm$ 0.0078     & 7.185 $\pm$ 4.837  & 0.9391$\pm$ 0.0565   & 0.5419   \\
\br
\end{tabular}
\caption{Sweep through the number of transients observed for a simulation that selects 3 fixed oscillators and adds a random Gaussian perturbation to their phases with standard deviation $10$.}
\label{tab:3fixed_10}
\end{table}

\begin{table}[h!]
\footnotesize
\begin{tabular}{lllllll}
\br
Param   & Value  & \begin{tabular}{@{}l@{}} Normalized \\ difference in area \end{tabular} & \begin{tabular}{@{}l@{}}Mean absolute\\ deviation\end{tabular} & Error rate           & \begin{tabular}{@{}l@{}} Area under\\ ROC curve\end{tabular} & \begin{tabular}{@{}l@{}} Interval \\width $(>90\%)$  \end{tabular}\\
\mr
$N_{pert}$ & 1     & 0.891 $\pm$ 0.1231  & 0.1058 $\pm$ 0.0706     & 46.889 $\pm$ 8.444  & 0.3337$\pm$ 0.1156   & 0.0299                     \\
$N_{pert}$ & 2 & 0.9214 $\pm$ 0.088 & 0.3684 $\pm$ 0.0897     & 50.444 $\pm$ 7.031  & 0.2938$\pm$ 0.0783   & 0.0474                     \\
$N_{pert}$ & 5     & 0.2075 $\pm$ 0.0319 & 0.0313 $\pm$ 0.0179     & 23.704 $\pm$ 10.379 & 0.7769$\pm$ 0.0874   & 0.2803                     \\
$N_{pert}$ & 10    & 0.0294 $\pm$ 0.0061 & 0.0033 $\pm$ 0.0014     & 1.926 $\pm$ 2.592   & 0.9894$\pm$ 0.0157   & 0.3855                     \\
$N_{pert}$ & 20    & 0.0184 $\pm$ 0.003  & 0.0015 $\pm$ 0.0007     & 0.074 $\pm$ 0.406   & 0.9999$\pm$ 0.0008   & 0.522                      \\
$N_{pert}$ & 40    & 0.0116 $\pm$ 0.0028 & 0.0015 $\pm$ 0.0005     & 0.0 $\pm$ 0.0         & 1.0$\pm$ 0.0         & 0.6206  \\
\br
\end{tabular}
\caption{Sweep through the number of transients observed for a simulation that selects 3 random oscillators and resets their phases to $0$.}
\label{tab:3rand_set0}
\end{table}

 \begin{table}[h!]
  \footnotesize
\begin{tabular}{lllllll}
\br
Param   & Value  & \begin{tabular}{@{}l@{}} Normalized \\ difference in area \end{tabular} & \begin{tabular}{@{}l@{}}Mean absolute\\ deviation\end{tabular} & Error rate           & \begin{tabular}{@{}l@{}} Area under\\ ROC curve\end{tabular} & \begin{tabular}{@{}l@{}} Interval \\width $(>90\%)$  \end{tabular}\\
\mr
$N_{res}$ & 1     & 0.8188 $\pm$ 0.1236 & 8633.13 $\pm$ 25993.8   & 45.852 $\pm$ 8.159 & 0.5239$\pm$ 0.0886   & 2012.5215                  \\
$N_{res}$ & 2     & 0.8895 $\pm$ 0.2663 & 3481.08 $\pm$ 9497.15 & 23.037 $\pm$ 11.635 & 0.7415$\pm$ 0.1298   & 1.6593                     \\
$N_{res}$ & 5     & 0.2267 $\pm$ 0.3482 & 2.5746 $\pm$ 11.3136      & 2.148 $\pm$ 7.063  & 0.9756$\pm$ 0.0733   & 0.75                       \\
$N_{res}$ & 10    & 0.0238 $\pm$ 0.0242 & 0.0061 $\pm$ 0.0078       & 0.074 $\pm$ 0.406  & 0.9979$\pm$ 0.0114   & 0.8436                     \\
$N_{res}$ & 20    & 0.0099 $\pm$ 0.0037 & 0.0024 $\pm$ 0.001        & 0.0 $\pm$ 0.0        & 1.0$\pm$ 0.0         & 0.9131                     \\
$N_{res}$ & 40    & 0.0077 $\pm$ 0.0033 & 0.0023 $\pm$ 0.001        & 0.0 $\pm$ 0.0        & 1.0$\pm$ 0.0         & 0.9329  \\
\br
\end{tabular}
\caption{Sweep through the number of transients observed for a simulation that carries out the optimization for a linear system as proposed in \cite{Pik2018}.}
\label{tab:pikovsky}
\end{table}

\clearpage

\newcommand{\newblock}{}
\bibliographystyle{iopart-num}
\bibliography{main}

\end{document}